# Design of thermo-piezoelectric microstructured bending actuators via multi-field asymptotic homogenization

Francesca Fantoni[1,*], Andrea Bacigalupo[2,*], Marco Paggi[2]

[1] DICATAM, Università degli Studi di Brescia, via Branze 43, 25123, Brescia, Italy
[2] IMT School for Advanced Studies Lucca, Piazza S.Francesco 19, 55100 Lucca, Italy

March 14, 2018

**Abstract**

The use of integrated MicroElectroMechanical systems (MEMS) is recently spread thanks to their improved sensitivity, accuracy and reliability. Accurate preliminary computations born from the need of high precision in the manufacturing process of such devices. Piezoelectric materials are broadly employed in this field as direct converters between mechanical and electrical signals and some of these piezoelectric materials show pyroelectric features, which involve thermo-electrical interactions. Pyroelectric bending actuators are analyzed in the present study in plane conditions. They consists of active PZT layers with in-plane polarization and a microstructured composite layer characterized by a periodic microstructure where PZT fibers with an out of plane polarization are immersed in a polymeric matrix. The constitutive law of the composite layer at the macroscale has been determined by means of a multi-field asymptotic homogenization technique, recently developed for thermo-piezoelectric materials. Overall constitutive equations characterizing the behavior of the microstructured layer at the macroscale have been derived and the closed form of the overall constitutive tensors has been provided for the equivalent first-order (Cauchy) homogenized continuum. Deflection of unimorph and bimorph bender actuators has been investigated in relation to their geometrical features, exploiting the out of plane piezoelectric properties of the composite layer, which modify the stiffness of the entire bender. An accurate description of benders behavior at the structural length scale is of fundamental importance in order to design devices with high performances. In this regard, the influence of the microstructure on the global response of the actuator is investigated in the present study in order to understand how the composite material can be tailored to meet specific design requirements.

## 1 Introduction

Piezoelecricity is the property of some crystalline materials to generate an electric field if mechanically deformed (direct effect). Such deformation of the crystalline structure, which becomes electrically polarized, is completely reversible. The inverse piezoelectric effect on the contrary consists of experiencing mechanical deformations in the presence of an electric field (Yang, 2004). By virtue of these two principles, piezoelectric materials can be used to create sensors (direct effect) and actuators (inverse effect). Since the beginning of the twentieth century, scientific research related to piezoelectric materials has grown steadily (Brand et al., 2015; Kim et al., 2011; Anton and Sodano, 2007), fueled by the interest to fulfill industry needs, involving numerous engineering sectors, from electronic components (transformers, frequency generators) to telecommunications, from the automotive field (injection systems, vibration control) to the biomedical field.
Some piezoelectric materials are also pyroelectric. Pyroelectricity involves thermo-electrical interactions, being the polarization of the crystalline material dependent in this case on temperature (Moulson and Herbert, 2003; Batra and Aggarwal, 2013). For pyroelectric materials the nature of thermal gradient modifies the direction of the pyroelectric current and as a result of this change in temperature, the material becomes polarized establishing an electrical potential.

---

[*]Corresponding authors: Tel:+39 0583 4326613,
E-mail addresses: francesca.fantoni@unibs.it; andrea.bacigalupo@imtlucca.it

Restricting the attention on actuators, the most common piezo/pyroelectric devices are of two types: multilayer (or stack) actuators and benders. Stack actuators are created by layering multiple piezoelectric elements, each enclosed between two electrodes, taking advantage of their combined expansion in order to produce movement and force. Bender actuators, on the other hand, refer to two basic models: unimorph and bimorph. Unimorph bender consists of a piezoelectric layer and a passive one. Bimorph bender is characterized by a central passive layer sandwiched between two piezoelectric laminae. Electrodes are applied on the upper and lower surfaces of each piezoelectric layer (Kang and Wang, 2010; Yan et al., 2011). For this type of actuator, deformations induced by the electric field lead to deflection of the device which act as a cantilever.

Piezo/pyroelectric actuators are used for example as signal sources, micro and nano-positioning systems (Okazaki, 1990), micro-mirrors (Cheng et al., 2001), micro-grippers (Wang et al., 1999), vibration damping (Hagood and von Flotow, 1991) and noise control. Piezoelectric laminae are often made of Lead Zirconate Titanate ($P_b(Zr_xTi_{1-x})O_3$), commonly called PZT, which manifests excellent piezoelectric properties, including rapid signal response, large electro-mechanical coupling factor, physical and chemical stability, low energy consumption, and miniaturization potential.

Miniaturized and integrated microelectromechanical systems (MEMS) actuators and sensors are increasingly developed thanks to the recent advances in microprocessing technologies (Saadon and Sidek, 2011; Gardner et al., 2001). This leads to high precision manufacturing needs, with a remarkable demand of preliminary computations (Heinonen et al., 2007) which, in many cases, become cumbersome and time consuming.

In this regard, homogenization methods can be conveniently exploited in order to describe the effects of the microstructure on the overall behavior of the material in a rigorous and synthetic way, avoiding the challenging need to perform a numerical computation of the whole heterogeneous medium. Homogenization techniques therefore reveal to be an useful and convenient method to take into account the microstructural heterogeneity, especially in the case of periodic composite media, such as the above-mentioned multi-layered actuators.

The overall static and dynamic properties of periodic composite materials can generally be determined by asymptotic homogenization methods (Fantoni et al., 2017; Bacigalupo et al., 2016a,b; Bakhvalov and Panasenko, 1984; Bensoussan et al., 1978), variational-asymptotic techniques (Bacigalupo et al., 2014; Bacigalupo, 2014; Peerlings and Fleck, 2004; Smyshlyaev and Cherednichenko, 2000), and computational homogenization techniques (Trovalusci et al., 2017; De Bellis and Bacigalupo, 2017; Bacigalupo et al., 2017; Addessi et al., 2016, 2013; Bacca et al., 2013a,b,c; Bigoni and Drugan, 2007; Forest and Sab, 1998; Forest and Trinh, 2011; Miehe et al., 1999; Kouznetsova et al., 2004). The present study aims at providing a characterization of the deflection of multi-layered bending actuators, in particular unimorh and bimorph benders. These lasts are considered made of one or two active piezoelectric laminae and a composite layer, which replaces the usual passive layer, characterized by a periodic microstructure. In the present study, the multi-filed asymptotic homogenization technique recently developed in (Fantoni et al., 2017) for periodic thermo-piezoelectric materials, is exploited to describe the constitutive law of the microstructured composite layer at the macroscale. In particular, the exact closed forms of the overall thermo-piezoelectric tensors of the composite layer are determined for the case of plane stress and plane strain conditions, thus avoiding the need to model the whole microstructured heterogeneous stratum. Field equations for the homogenized first-order thermo-piezoelectric medium are derived, following the procedure developed in (Bacigalupo and Gambarotta, 2014; Smyshlyaev and Cherednichenko, 2000; Bakhvalov and Panasenko, 1984) for composite elastic media with periodic microstructure and particularized by (Fantoni et al., 2017) to the case of thermo-piezoelectric materials. Analyzed benders remain heterogeneous at the structural length scale being composed by the piezolectric laminae and the homogenized composite layer.

The paper is organized as follows. In *Section 2* the geometry of a generic periodic thermo-piezoelectric composite material is illustrated and the corresponding constitutive equations and balance relations are introduced. The developed multi-field asymptotic homogenization technique is described in *Section 3*, based on down-scaling relations correlating the microscopic fields to the macroscopic ones. Periodic perturbation functions, which take into account the effects of the microstructural heterogeneity on the micro-fields, are derived from the solution of non homogeneous recursive cell problems defined over the unit periodic cell. In the same *Section*, average field equations of infinite order have been obtained and the closed form of the overall constitutive tensors has been determined for the equivalent first-order homogenized continuum. These results are used in *Section 4* to derive overall tensors of a composite thermo-piezoelectric material with periodic microstructure characterized by circular or square inclusions. Deflection of unimorph and bimorph bender actuators has been investigated varying the geometric features of the benders, loaded by imposed electric potential and relative temperature gradients. Finally, conclusions are reported in *Section 5*.

## 2 Multiscale modeling of periodic thermo-piezoelectric composites

The present multiscale description deals with an heterogeneous composite material characterized by a periodic microstructure, as the one depicted in figure 1, under the assumption of small strains. The continuum is described as a linear thermo-piezoelectric Cauchy medium (Mindlin, 1974), subject to volume forces represented by body forces, free charge densities, and temperature changes.

Position vector $\mathbf{x} = x_1\,\mathbf{e}_1 + x_2\,\mathbf{e}_2$ identifies the position of each material point in the continuum, with reference to the orthogonal coordinates system with origin at point $O$ and base $\{\mathbf{e_1}, \mathbf{e_2}\}$ (see figure 1-(a)). Without losing generality, the notation has been restricted to the two dimensional case, for the sake of simplicity. The entire periodic medium can be obtained by spanning a periodic cell $\mathcal{A} = [0,\varepsilon] \times [0,\delta\varepsilon]$ by the two orthogonal vectors $\mathbf{v}_1 = d_1\,\mathbf{e}_1 = \varepsilon\mathbf{e}_1$ and $\mathbf{v}_2 = d_2\,\mathbf{e}_2 = \delta\varepsilon\,\mathbf{e}_2$, with $\varepsilon$ the characteristic size of the periodic cell $\mathcal{A}$. The micro constitutive tensors, namely micro-elasticity tensor $\mathbb{C}^{(m,\varepsilon)}$, the micro-dielectric permittivity tensor (at constant strain) $\boldsymbol{\beta}^{(m,\varepsilon)}$, the micro-heat conduction tensor $\boldsymbol{K}^{(m,\varepsilon)}$, the micro piezoelectric stress/charge tensor $\boldsymbol{e}^{(m,\varepsilon)}$, the micro-thermal dilatation tensor $\boldsymbol{\alpha}^{(m,\varepsilon)}$, and the micro pyrolelectric vector $\boldsymbol{\gamma}^{(m,\varepsilon)}$, obey the following properties because of the $\mathcal{A}$ periodicity of the material

$$\mathbb{C}^{(m,\varepsilon)}(\mathbf{x}+\mathbf{v}_i) = \mathbb{C}^{(m,\varepsilon)}(\mathbf{x}), \quad i=1,2, \quad \forall \mathbf{x} \in \mathcal{A}, \tag{1a}$$

$$\boldsymbol{\beta}^{(m,\varepsilon)}(\mathbf{x}+\mathbf{v}_i) = \boldsymbol{\beta}^{(m,\varepsilon)}(\mathbf{x}), \quad i=1,2, \quad \forall \mathbf{x} \in \mathcal{A}, \tag{1b}$$

$$\boldsymbol{K}^{(m,\varepsilon)}(\mathbf{x}+\mathbf{v}_i) = \boldsymbol{K}^{(m,\varepsilon)}(\mathbf{x}), \quad i=1,2, \quad \forall \mathbf{x} \in \mathcal{A}, \tag{1c}$$

$$\boldsymbol{e}^{(m,\varepsilon)}(\mathbf{x}+\mathbf{v}_i) = \boldsymbol{e}^{(m,\varepsilon)}(\mathbf{x}), \quad i=1,2, \quad \forall \mathbf{x} \in \mathcal{A}, \tag{1d}$$

$$\boldsymbol{\alpha}^{(m,\varepsilon)}(\mathbf{x}+\mathbf{v}_i) = \boldsymbol{\alpha}^{(m,\varepsilon)}(\mathbf{x}), \quad i=1,2, \quad \forall \mathbf{x} \in \mathcal{A}, \tag{1e}$$

$$\boldsymbol{\gamma}^{(m,\varepsilon)}(\mathbf{x}+\mathbf{v}_i) = \boldsymbol{\gamma}^{(m,\varepsilon)}(\mathbf{x}), \quad i=1,2, \quad \forall \mathbf{x} \in \mathcal{A}. \tag{1f}$$

A non dimensional unit cell $\mathcal{Q} = [0,1] \times [0,\delta]$ can be conveniently introduced to reproduce the periodic microstructure, rescaling the periodic cell $\mathcal{A}$ by the characteristic length $\varepsilon$. Two variables, the macroscopic (slow) one $\mathbf{x} \in \mathcal{A}$ and the microscopic (fast) one $\boldsymbol{\xi} = \mathbf{x}/\varepsilon \in \mathcal{Q}$, allow the separation of the macro and the micro scales (Bacigalupo, 2014; Smyshlyaev, 2009; Peerlings and Fleck, 2004; Bakhvalov and Panasenko, 1984). Therefore, constitutive tensors (1a)-(1f) can be defined over $\mathcal{Q}$ in the following form, dependent only on the microscopic variable $\boldsymbol{\xi}$

$$\begin{aligned}\mathbb{C}^{(m,\varepsilon)}(\mathbf{x}) &= \mathbb{C}^m(\boldsymbol{\xi}=\mathbf{x}/\varepsilon), \quad \boldsymbol{\beta}^{(m,\varepsilon)}(\mathbf{x}) = \boldsymbol{\beta}^m(\boldsymbol{\xi}=\mathbf{x}/\varepsilon), \quad \boldsymbol{K}^{(m,\varepsilon)}(\mathbf{x}) = \boldsymbol{K}^m(\boldsymbol{\xi}=\mathbf{x}/\varepsilon),\\ \boldsymbol{e}^{(m,\varepsilon)}(\mathbf{x}) &= \boldsymbol{e}^m(\boldsymbol{\xi}=\mathbf{x}/\varepsilon), \quad \boldsymbol{\alpha}^{(m,\varepsilon)}(\mathbf{x}) = \boldsymbol{\alpha}^m(\boldsymbol{\xi}=\mathbf{x}/\varepsilon), \quad \boldsymbol{\gamma}^{(m,\varepsilon)}(\mathbf{x}) = \boldsymbol{\gamma}^m(\boldsymbol{\xi}=\mathbf{x}/\varepsilon).\end{aligned} \tag{2}$$

The micro stress $\boldsymbol{\sigma}(\mathbf{x})$, the micro electric displacement $\mathbf{D}(\mathbf{x})$, and the micro heat flux $\mathbf{q}(\mathbf{x})$ are determined respectively by the following constitutive relations (Mindlin, 1974)

$$\boldsymbol{\sigma}(\mathbf{x}) = \mathbb{C}^{(m,\varepsilon)}(\mathbf{x})\,\boldsymbol{\varepsilon}(\mathbf{x}) + \boldsymbol{e}^{(m,\varepsilon)}(\mathbf{x})\,\nabla\phi(\mathbf{x}) - \boldsymbol{\alpha}^{(m,\varepsilon)}(\mathbf{x})\,\theta(\mathbf{x}), \tag{3a}$$

$$\mathbf{D}(\mathbf{x}) = \tilde{\boldsymbol{e}}^{(m,\varepsilon)}(\mathbf{x})\,\boldsymbol{\varepsilon}(\mathbf{x}) - \boldsymbol{\beta}^{(m,\varepsilon)}(\mathbf{x})\,\nabla\phi(\mathbf{x}) + \boldsymbol{\gamma}^{(m,\varepsilon)}(\mathbf{x})\,\theta(\mathbf{x}), \tag{3b}$$

$$\mathbf{q}(\mathbf{x}) = -\boldsymbol{K}^{(m,\varepsilon)}(\mathbf{x})\,\nabla\theta(\mathbf{x}), \tag{3c}$$

where $\tilde{e}^m_{ijk} = e^m_{jki}$ and the time derivative of $\mathbf{u}(\mathbf{x})$ and $\theta(\mathbf{x})$ has been neglected in equation (3c) in the hypothesis of quasi-static processes (Mindlin, 1974; Nowacki, 1986). The following local balance equations are satisfied respectively by the stress field $\boldsymbol{\sigma}(x)$, the electric displacement field $\mathbf{D}(\mathbf{x})$, and the heat flux $\mathbf{q}(x)$

$$\nabla \cdot \boldsymbol{\sigma}(\mathbf{x}) + \mathbf{b}(\mathbf{x}) = \mathbf{0}, \quad \nabla \cdot \mathbf{D}(\mathbf{x}) - \rho_e(\mathbf{x}) = 0, \quad \nabla \cdot \boldsymbol{q}(\mathbf{x}) + r(\mathbf{x}) = 0, \tag{4}$$

where volume forces are represented by body forces $\mathbf{b}(\mathbf{x})$, free charge densities $\rho_e(\mathbf{x})$, and heat sources $r(\mathbf{x})$, all dependent only on the slow variable $\mathbf{x}$. Volume forces are assumed to be $\mathcal{L}$-periodic, with vanishing mean values on $\mathcal{L}$, being $\mathcal{L} = [0,L] \times [0,\delta L]$. $\mathcal{L}$ can be considered as a true representative portion of the whole body, because, for the scales separation condition, the structural length $L$ is much greater than the microstructural one $\varepsilon$ ($L >> \varepsilon$).

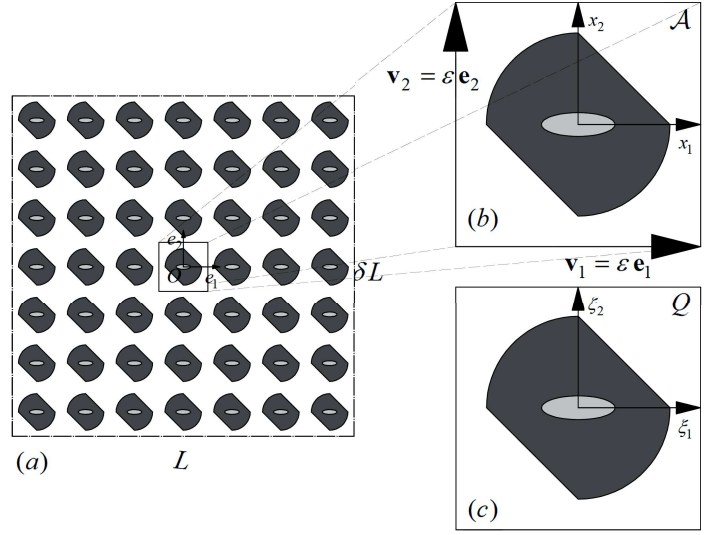

Figure 1: *(a) Periodic microstructure of the heterogeneous medium with structural characteristic size $L$; (b) Periodic cell $\mathcal{A}$ with microstructural characteristic size $\varepsilon$ and periodicity vectors $\mathbf{v}_1$ and $\mathbf{v}_2$; (c) Periodic unit cell $\mathcal{Q}$.*

Substituting constitutive equations (3a)-(3c) into balance equations (4), one obtains the following set of partial differential equations:

$$\nabla \cdot \left(\mathbb{C}^{(m,\varepsilon)}(\mathbf{x}) \nabla \mathbf{u}(\mathbf{x})\right) + \nabla \cdot \left(\mathbf{e}^{(m,\varepsilon)}(\mathbf{x}) \nabla \phi(\mathbf{x})\right) - \nabla \cdot \left(\boldsymbol{\alpha}^{(m,\varepsilon)}(\mathbf{x}) \theta(\mathbf{x})\right) + \mathbf{b}(\mathbf{x}) = \mathbf{0}, \tag{5a}$$

$$\nabla \cdot \left(\tilde{\mathbf{e}}^{(m,\varepsilon)}(\mathbf{x}) \nabla \mathbf{u}(\mathbf{x})\right) - \nabla \cdot \left(\boldsymbol{\beta}^{(m,\varepsilon)}(\mathbf{x}) \nabla \phi(\mathbf{x})\right) + \nabla \cdot \left(\boldsymbol{\gamma}^{(m,\varepsilon)}(\mathbf{x}) \theta(\mathbf{x})\right) - \rho_e(\mathbf{x}) = 0, \tag{5b}$$

$$\nabla \cdot \left(\boldsymbol{K}^{(m,\varepsilon)}(\mathbf{x}) \nabla \theta(\mathbf{x})\right) + r(\mathbf{x}) = 0. \tag{5c}$$

The following continuity conditions hold for a perfectly bounded interface, with $[[f]] = f^i(\Sigma) - f^j(\Sigma)$ denoting the jump of the values of function $f$ at the interface $\Sigma$ between two different phases $i$ and $j$ in the periodic cell $\mathcal{A}$

$$[[\mathbf{u}(\mathbf{x})]]|_{\boldsymbol{x} \in \Sigma} = \mathbf{0}, \quad \left[\left[\left(\mathbb{C}^{(m,\varepsilon)}(\mathbf{x}) \nabla \mathbf{u}(\mathbf{x}) + \mathbf{e}^{(m,\varepsilon)}(\mathbf{x}) \nabla \phi(\mathbf{x}) - \boldsymbol{\alpha}^{(m,\varepsilon)}(\mathbf{x}) \theta(\mathbf{x})\right) \cdot \mathbf{n}\right]\right]\Big|_{\boldsymbol{x} \in \Sigma} = \mathbf{0}, \tag{6a}$$

$$[[\phi(\mathbf{x})]]|_{\boldsymbol{x} \in \Sigma} = \mathbf{0}, \quad \left[\left[\left(\tilde{\mathbf{e}}^{(m,\varepsilon)}(\mathbf{x}) \nabla \mathbf{u}(\mathbf{x}) - \boldsymbol{\beta}^{(m,\varepsilon)}(\mathbf{x}) \nabla \phi(\mathbf{x}) + \boldsymbol{\gamma}^{(m,\varepsilon)}(\mathbf{x}) \theta(\mathbf{x})\right) \cdot \mathbf{n}\right]\right]\Big|_{\boldsymbol{x} \in \Sigma} = \mathbf{0}, \tag{6b}$$

$$[[\theta(\mathbf{x})]]|_{\boldsymbol{x} \in \Sigma} = \mathbf{0}, \quad \left[\left[\boldsymbol{K}^{(m,\varepsilon)}(\mathbf{x}) \nabla \theta(\mathbf{x}) \cdot \mathbf{n}\right]\right]\Big|_{\boldsymbol{x} \in \Sigma} = \mathbf{0}, \tag{6c}$$

where $\mathbf{n}$ denotes the outward normal to the interface $\Sigma$.

The micro fields depend on both the slow variable $\mathbf{x}$ and the fast one $\boldsymbol{\xi}$, because of the $\mathcal{Q}$-periodicity of the micro constitutive tensors (2) and the $\mathcal{L}$-periodicity of volume forces

$$\mathbf{u} = \mathbf{u}\left(\mathbf{x}; \boldsymbol{\xi} = \frac{\mathbf{x}}{\varepsilon}\right), \quad \phi = \phi\left(\mathbf{x}; \boldsymbol{\xi} = \frac{\mathbf{x}}{\varepsilon}\right), \quad \theta = \theta\left(\mathbf{x}; \boldsymbol{\xi} = \frac{\mathbf{x}}{\varepsilon}\right).$$

$\mathcal{Q}$-periodicity of coefficients of equations (5a)-(5c), both analytically and numerically. Therefore, it is convenient to replace the heterogeneous model with an equivalent homogeneous one, obtaining in this way equations whose coefficients are not rapidly oscillating. In what follows, the formulation of an equivalent first-order thermo-piezoelectric continuum will be presented, as derived in (Fantoni et al., 2017), and the exact expressions of the overall constitutive tensors will be provided. In the equivalent homogenized medium, the macro fields for each material points $\mathbf{x}$ are the displacement field $\mathbf{U}(\mathbf{x}) = U_i \mathbf{e}_i$, the electric potential $\Phi(\mathbf{x})$, and the relative temperature $\Theta(\mathbf{x})$. They are $\mathcal{L}$-periodic functions dependent only on the slow variable $\mathbf{x}$.

## 3 Asymptotic homogenization approach to thermo-piezoelectric composites

The following down-scaling relations can be derived for the three micro-fields, representing these lasts trough an asymptotic expansions in powers of the microstructural length scale $\varepsilon$ (Bakhvalov and Panasenko, 1984; Bensoussan et al., 1978)

$$
\begin{aligned}
u_k(\mathbf{x}; \boldsymbol{\xi}) =& U_k(\mathbf{x}) + \varepsilon \left( N_{kpq_1}^{(1)}(\boldsymbol{\xi}) \frac{\partial U_p(\mathbf{x})}{\partial x_{q_1}} + \tilde{N}_{kq_1}^{(1)}(\boldsymbol{\xi}) \frac{\partial \Phi(\mathbf{x})}{\partial x_{q_1}} + \hat{N}_k^{(1)}\Theta(x) \right)\bigg|_{\boldsymbol{\xi}=\mathbf{x}/\varepsilon} \\
&+ \varepsilon^2 \left( N_{kpq_1q_2}^{(2)}(\boldsymbol{\xi}) \frac{\partial^2 U_p(\mathbf{x})}{\partial x_{q_1}\partial x_{q_2}} + \tilde{N}_{kq_1q_2}^{(2)}(\boldsymbol{\xi}) \frac{\partial^2 \Phi(\mathbf{x})}{\partial x_{q_1}\partial x_{q_2}} + \hat{N}_{kq_1}^{(2)} \frac{\partial \Theta(\mathbf{x})}{\partial x_{q_1}} \right)\bigg|_{\boldsymbol{\xi}=\mathbf{x}/\varepsilon} + \mathcal{O}(\varepsilon)^3,
\end{aligned} \quad (7a)
$$

$$
\begin{aligned}
\phi(\mathbf{x}; \boldsymbol{\xi}) =& \phi(\mathbf{x}) + \varepsilon \left( W_{q_1}^{(1)}(\boldsymbol{\xi}) \frac{\partial \Phi(\mathbf{x})}{\partial x_{q_1}} + \tilde{W}_{pq_1}^{(1)}(\boldsymbol{\xi}) \frac{\partial U_p(\mathbf{x})}{\partial x_{q_1}} + \hat{W}^{(1)}(\boldsymbol{\xi})\Theta(\mathbf{x}) \right)\bigg|_{\boldsymbol{\xi}=\mathbf{x}/\varepsilon} \\
&+ \varepsilon^2 \left( W_{q_1q_2}^{(2)}(\boldsymbol{\xi}) \frac{\partial^2 \Phi(\mathbf{x})}{\partial x_{q_1}\partial x_{q_2}} + \tilde{W}_{pq_1q_2}^{(2)}(\boldsymbol{\xi}) \frac{\partial^2 U_p(\mathbf{x})}{\partial x_{q_1}\partial x_{q_2}} + \hat{W}_{q_1}^{(2)}(\boldsymbol{\xi}) \frac{\partial \Theta(\mathbf{x})}{\partial x_{q_1}} \right)\bigg|_{\boldsymbol{\xi}=\mathbf{x}/\varepsilon} + \mathcal{O}(\varepsilon)^3,
\end{aligned} \quad (7b)
$$

$$
\theta(\mathbf{x}; \boldsymbol{\xi}) = \Theta(\mathbf{x}) + \varepsilon \left( M_{q_1}^{(1)}(\boldsymbol{\xi}) \frac{\partial \Theta(\mathbf{x})}{\partial x_{q_1}} \right)\bigg|_{\boldsymbol{\xi}=\mathbf{x}/\varepsilon} + \varepsilon^2 \left( M_{q_1q_2}^{(2)}(\boldsymbol{\xi}) \frac{\partial^2 \Theta(\mathbf{x})}{\partial x_{q_1}\partial x_{q_2}} \right)\bigg|_{\boldsymbol{\xi}=\mathbf{x}/\varepsilon} + \mathcal{O}(\varepsilon)^3. \quad (7c)
$$

Relations (7a)-(7c) have been proved in detail in (Fantoni et al., 2017). In equations (7a)-(7c), $N_{kpq_1}^{(1)}$, $\tilde{N}_{kq_1}^{(1)}$, $\hat{N}_k^{(1)}$, $N_{kpq_1q_2}^{(2)}$, $\tilde{N}_{kq_1q_2}^{(2)}$, $\hat{N}_{kq_1}^{(2)}$ are the perturbation functions for the micro displacement field, $W_{q_1}^{(1)}$, $\tilde{W}_{pq_1}^{(1)}$, $\hat{W}^{(1)}$, $W_{q_1q_2}^{(2)}$, $\tilde{W}_{pq_1q_2}^{(2)}$, $\hat{W}_{q_1}^{(2)}$ are the perturbation functions for the micro electric potential field, and finally $M_{q_1}^{(1)}$, $M_{q_1q_2}^{(2)}$ are the perturbation functions for the micro relative temperature field. They express the influence of the fast variable $\boldsymbol{\xi} = \mathbf{x}/\varepsilon$ on the micro fields reflecting the effects of material inhomogeneities on the micro fields. They are $\mathcal{Q}$-periodic and it is assumed that they have vanishing mean values over the unit cell $\mathcal{Q}$. Therefore they all satisfy the following normalization condition over $\mathcal{Q}$

$$
\langle (\cdot) \rangle = \frac{1}{\delta} \int_{\mathcal{Q}} (\cdot) \, d\boldsymbol{\xi} \quad (8)
$$

Substituting down scaling relations (7a)-(7c) into field equations (5a)-(5c) asymptotically expanded, a series of recursive differential problems can be obtained. Field equations of the equivalent first-order thermo-piezoelectric continuum can be obtained considering only the $\varepsilon^{-1}$ and $\varepsilon^0$ terms of the sequence of PDEs derived by the asymptotic procedure, according to (Fantoni et al., 2017), exploiting the property

$$
\frac{\partial}{\partial x_j} f(\mathbf{x}; \boldsymbol{\xi} = \frac{\mathbf{x}}{\varepsilon}) = \left( \frac{\partial f}{\partial x_j} + \frac{1}{\varepsilon}\frac{\partial f}{\partial \xi_j} \right)\bigg|_{\boldsymbol{\xi}=\frac{\mathbf{x}}{\varepsilon}} = \left( \frac{\partial f}{\partial x_j} + \frac{f_j}{\varepsilon} \right)\bigg|_{\boldsymbol{\xi}=\frac{\mathbf{x}}{\varepsilon}}.
$$

In order to obtain a set of PDEs with constant coefficients, the fluctuation functions must satisfy non homogeneous equations known as cell problems. The mechanical and the electric problems remain coupled in the asymptotically expanded microscale field equations, resulting in a strong coupling between the micro-displacement field and the micro-electric potential one. At the order $\varepsilon^{-1}$ perturbation functions must satisfy the set of three cell problems presented below. $N_{kpq_1}^{(1)}$ and $\tilde{W}_{pq_1}^{(1)}$ are the solution of

$$
\begin{cases}
\left( C_{ijkl}^m N_{kpq_1,l}^{(1)} \right)_{,j} + \left( e_{ijk}^m \tilde{W}_{pq_1,k}^{(1)} \right)_{,j} + C_{ijpq_1,j}^m = n_{ipq_1}^{(1)} \\
\left( e_{kli}^m N_{kpq_1,l}^{(1)} \right)_{,i} - \left( \beta_{il}^m \tilde{W}_{pq_1,l}^{(1)} \right)_{,i} + e_{pq_1i,i}^m = \tilde{w}_{pq_1}^{(1)}
\end{cases}, \quad (9)
$$

with interface conditions defined at the interface $\Sigma_1$ between two different phases in the unit cell $\mathcal{Q}$

$$\left[\left[N^{(1)}_{kpq_1}\right]\right]\Big|_{\boldsymbol{\xi}\in\Sigma_1} = 0,$$
$$\left[\left[\tilde{W}^{(1)}_{pq_1}\right]\right]\Big|_{\boldsymbol{\xi}\in\Sigma_1} = 0,$$
$$\left[\left[\left(C^m_{ijkl}\left(\delta_{kp}\delta_{lq_1} + N^{(1)}_{kpq_1,l}\right) + e^m_{ijk}\tilde{W}^{(1)}_{pq_1,k}\right)n_j\right]\right]\Big|_{\boldsymbol{\xi}\in\Sigma_1} = 0,$$
$$\left[\left[\left(e^m_{kli}\left(\delta_{kp}\delta_{lq_1} + N^{(1)}_{kpq_1,l}\right) - \beta^m_{il}\tilde{W}^{(1)}_{pq_1,l}\right)n_i\right]\right]\Big|_{\boldsymbol{\xi}\in\Sigma_1} = 0. \tag{10}$$

Perturbation functions $\tilde{N}^{(1)}_{kq_1}$ and $W^{(1)}_{q_1}$ are determined from the following cell problem

$$\begin{cases} \left(C^m_{ijkl}\tilde{N}^{(1)}_{kq_1,l}\right)_{,j} + \left(e^m_{ijk}W^{(1)}_{q_1,k}\right)_{,j} + e^m_{ijq_1,j} = \tilde{n}^{(1)}_{iq_1} \\ \left(e^m_{kli}\tilde{N}^{(1)}_{kq_1,l}\right)_{,i} - \left(\beta^m_{il}W^{(1)}_{q_1,l}\right)_{,i} - \beta^m_{iq_1,i} = w^{(1)}_{q_1} \end{cases}, \tag{11}$$

whose interface conditions are expressed as

$$\left[\left[\tilde{N}^{(1)}_{kq_1}\right]\right]\Big|_{\boldsymbol{\xi}\in\Sigma_1} = 0,$$
$$\left[\left[W^{(1)}_{q_1}\right]\right]\Big|_{\boldsymbol{\xi}\in\Sigma_1} = 0,$$
$$\left[\left[\left(C^m_{ijkl}\tilde{N}^{(1)}_{kq_1,l} + e^m_{ijk}\left(\delta_{kq_1} + W^{(1)}_{q_1,k}\right)\right)n_j\right]\right]\Big|_{\boldsymbol{\xi}\in\Sigma_1} = 0,$$
$$\left[\left[\left(e^m_{kli}\tilde{N}^{(1)}_{kq_1,l} - \beta^m_{il}\left(\delta_{lq_1} + W^{(1)}_{q_1,l}\right)\right)n_i\right]\right]\Big|_{\boldsymbol{\xi}\in\Sigma_1} = 0. \tag{12}$$

Finally, perturbation functions $\hat{N}^{(1)}_k$ and $\hat{W}^{(1)}$ are provided by the following cell problem at the order $\varepsilon^{-1}$

$$\begin{cases} \left(C^m_{ijkl}\hat{N}^{(1)}_{k,l}\right)_{,j} + \left(e^m_{ijk}\hat{W}^{(1)}_{,k}\right)_{,j} - \alpha^m_{ij,j} = \hat{n}^{(1)}_i \\ \left(e^m_{kli}\hat{N}^{(1)}_{k,l}\right)_{,i} - \left(\beta^m_{il}\hat{W}^{(1)}_{,l}\right)_{,i} + \gamma^m_{i,i} = \hat{w}^{(1)} \end{cases}, \tag{13}$$

with interface conditions

$$\left[\left[\hat{N}^{(1)}_k\right]\right]\Big|_{\boldsymbol{\xi}\in\Sigma_1} = 0,$$
$$\left[\left[\hat{W}^{(1)}\right]\right]\Big|_{\boldsymbol{\xi}\in\Sigma_1} = 0,$$
$$\left[\left[\left(C^m_{ijkl}\hat{N}^{(1)}_{k,l} + e^m_{ijk}\hat{W}^{(1)}_{,k} - \alpha^m_{ij}\right)n_j\right]\right]\Big|_{\boldsymbol{\xi}\in\Sigma_1} = 0,$$
$$\left[\left[\left(e^m_{kli}\hat{N}^{(1)}_{k,l} - \beta^m_{il}\hat{W}^{(1)}_{,l} + \gamma^m_i\right)n_i\right]\right]\Big|_{\boldsymbol{\xi}\in\Sigma_1} = 0. \tag{14}$$

Solvability condition of problems (9), (11), and (13) in the class of $\mathcal{Q}$-periodic functions (Bakhvalov and Panasenko, 1984), together with respective interface conditions (10), (12), and (14) implies that

$$n^{(1)}_{ipq_1} = \langle C^m_{ijpq_1,j}\rangle = 0, \quad \tilde{w}^{(1)}_{pq_1} = \langle e^m_{pq_1 i,i}\rangle = 0, \quad \tilde{n}^{(1)}_{iq_1} = \langle e^m_{ijq_1,j}\rangle = 0,$$
$$w^{(1)}_{q_1} = \langle -\beta^m_{iq_1,i}\rangle = 0, \quad \hat{n}^{(1)}_i = \langle -\alpha^m_{ij,j}\rangle = 0, \quad \hat{w}^{(1)} = \langle \gamma^m_{i,i}\rangle = 0.$$

For what regards the heat diffusion problem, the cell problem at the order $\varepsilon^{-1}$ takes the form

$$\left(K^m_{ij}M^{(1)}_{q_1,j}\right)_{,i} + K^m_{iq_1,i} = m^{(1)}_{q_1}, \tag{15}$$

with interface conditions expressed in terms of perturbation function $M^{(1)}_{q_1}$ as

$$\left[\left[M^{(1)}_{q_1}\right]\right]\Big|_{\boldsymbol{\xi}\in\Sigma_1} = 0,$$

$$\left[\left[K_{ij}^m \left(M_{q_1,j}^{(1)} + \delta_{jq_1}\right) n_i\right]\right]\bigg|_{\boldsymbol{\xi}\in\Sigma_1} = 0. \tag{16}$$

Due to the $\mathcal{Q}$-periodicity of components $K_{ij}^m$, one has $m_{q_1}^{(1)} = \left\langle K_{iq_1,i}^m\right\rangle = 0$ for the solvability of differential problem (15) with interface conditions (16). Once perturbation functions involved in cell problem (9), (11), (13), and (15) at the order $\varepsilon^{-1}$ are determined, one can solve the cell problems at the order $\varepsilon^0$, derived from equations (5a)-(5c). Specifically, the following cell problem, symmetrized with respect to indices $q_1$ and $q_2$, allows to derive perturbation functions $N_{kpq_1q_2}^{(2)}$ and $\tilde{W}_{pq_1q_2}^{(2)}$ for the piezoelectric problem

$$\begin{cases}
\left(C_{ijkl}^m N_{kpq_1q_2,l}^{(2)}\right)_{,j} + \left(e_{ijk}^m \tilde{W}_{pq_1q_2,k}^{(2)}\right)_{,j} + \frac{1}{2}\left[\left(C_{ijkq_2}^m N_{kpq_1}^{(1)}\right)_{,j} + C_{iq_1pq_2}^m + C_{iq_2kl}^m N_{kpq_1,l}^{(1)} + \right. \\
\left. + \left(e_{ijq_2}^m \tilde{W}_{pq_1}^{(1)}\right)_{,j} + e_{iq_2k}^m \tilde{W}_{pq_1,k}^{(1)} + \left(C_{ijkq_1}^m N_{kpq_2}^{(1)}\right)_{,j} + C_{iq_2pq_1}^m + C_{iq_1kl}^m N_{kpq_2,l}^{(1)} + \left(e_{ijq_1}^m \tilde{W}_{pq_2}^{(1)}\right)_{,j} + \right. \\
\left. + e_{iq_1k}^m \tilde{W}_{pq_2,k}^{(1)}\right] = n_{ipq_1q_2}^{(2)}, \\[6pt]
\left(e_{kli}^m N_{kpq_1q_2,l}^{(2)}\right)_{,i} - \left(\beta_{il}^m \tilde{W}_{pq_1q_2,l}^{(2)}\right)_{,i} + \frac{1}{2}\left[\left(e_{kq_2i}^m N_{kpq_1}^{(1)}\right)_{,i} + e_{klq_2}^m N_{kpq_1,l}^{(1)} + e_{pq_2q_1}^m + \right. \\
\left. - \left(\beta_{iq_2}^m \tilde{W}_{pq_1}^{(1)}\right)_{,i} - \beta_{q_2l}^m \tilde{W}_{pq_1,l}^{(1)} + \left(e_{kq_1i}^m N_{kpq_2}^{(1)}\right)_{,i} + e_{klq_1}^m N_{kpq_2,l}^{(1)} + e_{pq_1q_2}^m - \left(\beta_{iq_1}^m \tilde{W}_{pq_2}^{(1)}\right)_{,i} + \right. \\
\left. - \beta_{q_1l}^m \tilde{W}_{pq_2,l}^{(1)}\right] = \tilde{w}_{pq_1q_2}^{(2)}
\end{cases} \tag{17}$$

with interface conditions

$$\left[\left[N_{kpq_1q_2}^{(2)}\right]\right]\bigg|_{\boldsymbol{\xi}\in\Sigma_1} = 0,$$

$$\left[\left[\tilde{W}_{pq_1q_2}^{(2)}\right]\right]\bigg|_{\boldsymbol{\xi}\in\Sigma_1} = 0,$$

$$\left[\left[\left(C_{ijkl}^m N_{kpq_1q_2,l}^{(2)} + e_{ijk}^m \tilde{W}_{pq_1q_2,k}^{(2)} + \frac{1}{2}\left(C_{ijkq_2}^m N_{kpq_1}^{(1)} + C_{ijkq_1}^m N_{kpq_2}^{(1)} + e_{ijq_2}^m \tilde{W}_{pq_1}^{(1)} + e_{ijq_1}^m \tilde{W}_{pq_2}^{(1)}\right)\right) n_j\right]\right]\bigg|_{\boldsymbol{\xi}\in\Sigma_1} = 0,$$

$$\left[\left[\left(e_{kli}^m N_{kpq_1q_2,l}^{(2)} - \beta_{il}^m \tilde{W}_{pq_1q_2,l}^{(2)} + \frac{1}{2}\left(e_{kq_2i}^m N_{kpq_1}^{(1)} - \beta_{iq_2}^m \tilde{W}_{pq_1}^{(1)} + e_{kq_1i}^m N_{kpq_2}^{(1)} - \beta_{iq_1}^m \tilde{W}_{pq_2}^{(1)}\right)\right) n_i\right]\right]\bigg|_{\boldsymbol{\xi}\in\Sigma_1} = 0. \tag{18}$$

Perturbation functions $\tilde{N}_{kq_1q_2}^{(2)}$ and $W_{q_1q_2}^{(2)}$ are the solutions of the following cell problem, whose expression is reported, once again, in the symmetrized form with respect to $q_1$ and $q_2$ indices

$$\begin{cases}
\left(C_{ijkl}^m \tilde{N}_{kq_1q_2,l}^{(2)}\right)_{,j} + \left(e_{ijk}^m W_{q_1q_2,k}^{(2)}\right)_{,j} + \frac{1}{2}\left[\left(C_{ijkq_2}^m \tilde{N}_{kq_1}^{(1)}\right)_{,j} + C_{iq_2kl}^m \tilde{N}_{kq_1,l}^{(1)} + \left(e_{ijq_2}^m W_{q_1}^{(1)}\right)_{,j} + \right. \\
\left. + e_{iq_1q_2}^m + e_{iq_2k}^m W_{q_1,k}^{(1)} + \left(C_{ijkq_1}^m \tilde{N}_{kq_2}^{(1)}\right)_{,j} + C_{iq_1kl}^m \tilde{N}_{kq_2,l}^{(1)} + \left(e_{ijq_1}^m W_{q_2}^{(1)}\right)_{,j} + e_{iq_2q_1}^m + e_{iq_1k}^m W_{q_2,k}^{(1)}\right] = \\
= \tilde{n}_{iq_1q_2}^{(2)}
\end{cases} ,$$

$$\begin{cases}
\left(e_{kli}^m \tilde{N}_{kq_1q_2,l}^{(2)}\right)_{,i} - \left(\beta_{il}^m W_{q_1q_2,l}^{(2)}\right)_{,i} + \frac{1}{2}\left[\left(e_{kq_2i}^m \tilde{N}_{kq_1}^{(1)}\right)_{,i} + e_{klq_2}^m \tilde{N}_{kq_1,l}^{(1)} - \left(\beta_{iq_2}^m W_{q_1}^{(1)}\right)_{,i} - \beta_{q_1q_2}^m + \right. \\
\left. -\beta_{q_2l}^m W_{q_1,l}^{(1)} + \left(e_{kq_1i}^m \tilde{N}_{kq_2}^{(1)}\right)_{,i} + e_{klq_1}^m \tilde{N}_{kq_2,l}^{(1)} - \left(\beta_{iq_1}^m W_{q_2}^{(1)}\right)_{,i} - \beta_{q_2q_1}^m - \beta_{q_1l}^m W_{q_2,l}^{(1)}\right] = \\
= w_{q_1q_2}^{(2)}
\end{cases} \tag{19}$$

with interface conditions expressed as

$$\left[\left[\tilde{N}_{kq_1q_2}^{(2)}\right]\right]\bigg|_{\boldsymbol{\xi}\in\Sigma_1} = 0,$$

$$\left[\left[W_{q_1q_2}^{(2)}\right]\right]\bigg|_{\boldsymbol{\xi}\in\Sigma_1} = 0,$$

$$\left[\left[\left\{C_{ijkl}^m \tilde{N}_{kq_1q_2,l}^{(2)} + e_{ijk}^m W_{q_1q_2,k}^{(2)} + \frac{1}{2}\left(C_{ijkq_2}^m \tilde{N}_{kq_1}^{(1)} + e_{ijq_2}^m W_{q_1}^{(1)} + C_{ijkq_1}^m \tilde{N}_{kq_2}^{(1)} + e_{ijq_1}^m W_{q_2}^{(1)}\right)\right\}n_j\right]\right]\bigg|_{\boldsymbol{\xi}\in\Sigma_1} = 0,$$

$$\left[\left[\left\{e_{kli}^m \tilde{N}_{kq_1q_2,l}^{(2)} - \beta_{il}^m W_{q_1q_2,l}^{(2)} + \frac{1}{2}\left(e_{kq_2i}^m \tilde{N}_{kq_1}^{(1)} - \beta_{iq_2}^m W_{q_1}^{(1)} + e_{kq_1i}^m \tilde{N}_{kq_2}^{(1)} - \beta_{iq_1}^m W_{q_2}^{(1)}\right)\right\}n_i\right]\right]\bigg|_{\boldsymbol{\xi}\in\Sigma_1} = 0. \quad (20)$$

Finally, perturbation functions $\hat{N}_{kq_1}^{(2)}$ and $\hat{W}_{q_1}^{(2)}$ are provided by the solution of the cell problem

$$\begin{cases}
\left(C_{ijkl}^m \hat{N}_{kq_1,l}^{(2)}\right)_{,j} + \left(e_{ijk}^m \hat{W}_{q_1,k}^{(2)}\right)_{,j} + \left(C_{ijkq_1}^m \hat{N}_k^{(1)}\right)_{,j} + C_{iq_1kl}^m \hat{N}_{k,l}^{(1)} + \left(e_{ijq_1}^m \hat{W}^{(1)}\right)_{,j} + \\
+ e_{iq_1k}^m \hat{W}_{,k}^{(1)} - \left(\alpha_{ij}^m M_{q_1}^{(1)}\right)_{,j} - \alpha_{iq_1}^m = \hat{n}_{iq_1}^{(2)} \\
\left(e_{kli}^m \hat{N}_{kq_1,l}^{(2)}\right)_{,i} - \left(\beta_{il}^m \hat{W}_{q_1,l}^{(2)}\right)_{,i} + \left(e_{kq_1i}^m \hat{N}_k^{(1)}\right)_{,i} + e_{klq_1}^m \hat{N}_{k,l}^{(1)} - \left(\beta_{iq_1}^m \hat{W}^{(1)}\right)_{,i} + \\
-\beta_{q_1l}^m \hat{W}_{,l}^{(1)} + \left(\gamma_i^m M_{q_1}^{(1)}\right)_{,i} + \gamma_{q_1}^m = \hat{w}_{q_2}^{(2)}
\end{cases},$$
(21)

with interface conditions

$$\left[\left[\hat{N}_{kq_1}^{(2)}\right]\right]\bigg|_{\boldsymbol{\xi}\in\Sigma_1} = 0,$$

$$\left[\left[\hat{W}_{q_1}^{(2)}\right]\right]\bigg|_{\boldsymbol{\xi}\in\Sigma_1} = 0,$$

$$\left[\left[\left(C_{ijkl}^m \hat{N}_{kq_1,l}^{(2)} + e_{ijk}^m \hat{W}_{q_1,k}^{(2)} + C_{ijkq_1}^m \hat{N}_k^{(1)} + e_{ijq_1}^m \hat{W}^{(1)} - \alpha_{ij}^m M_{q_1}^{(1)}\right)n_j\right]\right]\bigg|_{\boldsymbol{\xi}\in\Sigma_1} = 0,$$

$$\left[\left[\left(e_{kli}^m \hat{N}_{kq_1,l}^{(2)} - \beta_{il}^m \hat{W}_{q_1,l}^{(2)} + e_{kq_1i}^m \hat{N}_k^{(1)} - \beta_{iq_1}^m \hat{W}^{(1)} + \gamma_i^m M_{q_1}^{(1)}\right)n_i\right]\right]\bigg|_{\boldsymbol{\xi}\in\Sigma_1} = 0. \quad (22)$$

It results

$$n_{ipq_1q_2}^{(2)} = \frac{1}{2}\left\langle C_{iq_1pq_2}^m + C_{iq_2kl}^m N_{kpq_1,l}^{(1)} + e_{iq_2k}^m \tilde{W}_{pq_1,k}^{(1)} + C_{iq_2pq_1}^m + C_{iq_1kl}^m N_{kpq_2,l}^{(1)} + e_{iq_1k}^m \tilde{W}_{pq_2,k}^{(1)}\right\rangle, \quad (23a)$$

$$\tilde{w}_{pq_1q_2}^{(2)} = \frac{1}{2}\left\langle e_{klq_2}^m N_{kpq_1,l}^{(1)} + e_{pq_2q_1}^m - \beta_{q_2l}^m \tilde{W}_{pq_1,l}^{(1)} + e_{klq_1}^m N_{kpq_2,l}^{(1)} + e_{pq_1q_2}^m - \beta_{q_1l}^m \tilde{W}_{pq_2,l}^{(1)}\right\rangle, \quad (23b)$$

$$\tilde{n}_{iq_1q_2}^{(2)} = \frac{1}{2}\left\langle C_{iq_2kl}^m \tilde{N}_{kq_1,l}^{(1)} + e_{iq_1q_2} + e_{iq_2k}^m W_{q_1,k}^{(1)} + C_{iq_1kl}^m \tilde{N}_{kq_2,l}^{(1)} + e_{iq_2q_1}^m + e_{iq_1k}^m W_{q_2,k}^{(1)}\right\rangle, \quad (23c)$$

$$w_{q_1q_2}^{(2)} = \frac{1}{2}\left\langle e_{klq_2}^m \tilde{N}_{kq_1,l}^{(1)} - \beta_{q_1q_2}^m - \beta_{q_2l}^m W_{q_1,l}^{(1)} + e_{klq_1}^m \tilde{N}_{kq_2,l}^{(1)} - \beta_{q_2q_1}^m - \beta_{q_1l}^m W_{q_2,l}^{(1)}\right\rangle, \quad (23d)$$

$$\hat{n}_{iq_1}^{(2)} = \left\langle C_{iq_1kl}^m \hat{N}_{k,l}^{(1)} + e_{iq_1k} \hat{W}_{,k}^{(1)} - \alpha_{iq_1}^{(m)}\right\rangle, \quad (23e)$$

$$\hat{w}_{q_2}^{(2)} = \left\langle e_{klq_1}^m \hat{N}_{k,l}^{(1)} - \beta_{q_1l}^m \hat{W}_{,l}^{(1)} + \gamma_{q_1}^m\right\rangle. \quad (23f)$$

For the heat diffusion problem, at the order $\varepsilon^0$ the obtained cell problem has the form

$$\left(K_{ij}^m M_{q_1q_2,j}^{(2)}\right)_{,i} + \frac{1}{2}\left[\left(K_{iq_2}^m M_{q_1}^{(1)}\right)_{,i} + K_{q_1q_2}^m + K_{q_2j}^m M_{q_1,j}^{(1)} + \left(K_{iq_2}^m M_{q_2}^{(1)}\right)_{,i} + K_{q_2q_1}^m + K_{q_1j}^m M_{q_2,j}^{(1)}\right] = m_{q_1q_2}^{(2)} \quad (24)$$

with interface conditions in terms of perturbation functions taking the following form

$$\left[\left[M_{q_1q_2}^{(2)}\right]\right]\bigg|_{\boldsymbol{\xi}\in\Sigma_1} = 0,$$

$$\left[\left[K_{ij}^m\left(M_{q_1q_2,j}^{(2)} + \frac{1}{2}\left(\delta_{jq_2} M_{q_1}^{(1)} + \delta_{jq_1} M_{q_2}^{(1)}\right)\right)n_i\right]\right]\bigg|_{\boldsymbol{\xi}\in\Sigma_1} = 0, \quad (25)$$

and term $m_{q_1q_2}^{(2)}$ having the following expression

$$m_{q_1q_2}^{(2)} = \frac{1}{2}\langle K_{q_1q_2}^m + K_{q_2j}^m M_{q_1,j}^{(1)} + K_{q_2q_1}^m + K_{q_1j}^m M_{q_2,j}^{(1)}\rangle \quad (26)$$

For the solvability conditions, the constants (23a)-(23f) and (26) are determined once again by imposing the non homogeneous terms in equation (17), (19), (21), and (24) possess vanishing mean values over the unit cell $\mathcal{Q}$. This implies the $\mathcal{Q}$-periodicity of perturbation functions and guarantees the continuity and regularity of the micro-fields $u_k, \phi$ and $\theta$. Once perturbation functions are known from the resolution of cell problems, the following averaged field equations can be derived, exploiting asymptotic expansions (7a)-(7c) and truncating these lasts at the order $\varepsilon^0$

$$n^{(2)}_{ipq_1q_2} \frac{\partial^2 U_p(\mathbf{x})}{\partial x_{q_1} \partial x_{q_2}} + \tilde{n}^{(2)}_{iq_1q_2} \frac{\partial^2 \Phi(\mathbf{x})}{\partial x_{q_1} \partial x_{q_2}} - \hat{n}^{(2)}_{iq_1} \frac{\partial \Theta(\mathbf{x})}{\partial x_{q_1}} + b_i(\mathbf{x}) + \mathcal{O}(\varepsilon) = 0, \tag{27a}$$

$$\tilde{w}^{(2)}_{pq_1q_2} \frac{\partial^2 U_p(\mathbf{x})}{\partial x_{q_1} \partial x_{q_2}} - w^{(2)}_{q_1q_2} \frac{\partial^2 \Phi(\mathbf{x})}{\partial x_{q_1} \partial x_{q_2}} + \hat{w}^{(2)}_{q_1} \frac{\partial \Theta(\mathbf{x})}{\partial x_{q_1}} - \rho_e(\mathbf{x}) + \mathcal{O}(\varepsilon) = 0, \tag{27b}$$

$$m^{(2)}_{q_1q_2} \frac{\partial^2 \Theta(\mathbf{x})}{\partial x_{q_1} \partial x_{q_2}} + r(\mathbf{x}) + \mathcal{O}(\varepsilon) = 0. \tag{27c}$$

The $\mathcal{L}$-periodic macroscopic solutions $\mathbf{U}(\mathbf{x}), \Phi(\mathbf{x})$, and $\Theta(\mathbf{x})$ of differential problems (27) are required to fulfill the following normalization conditions

$$\frac{1}{\delta L^2} \int_\mathcal{L} U_p(\mathbf{x}) \, d\mathbf{x} = 0, \quad \frac{1}{\delta L^2} \int_\mathcal{L} \Phi(\mathbf{x}) \, d\mathbf{x} = 0, \quad \frac{1}{\delta L^2} \int_\mathcal{L} \Theta(\mathbf{x}) \, d\mathbf{x} = 0. \tag{28}$$

Field equations (27) can be formally solved by performing an asymptotic expansion of the macro-fields in powers of $\varepsilon$:

$$U_k(\mathbf{x}) = \sum_{j=0}^{+\infty} \varepsilon^j \, U_k^{(j)}(\mathbf{x}), \tag{29a}$$

$$\Phi(\mathbf{x}) = \sum_{j=0}^{+\infty} \varepsilon^j \, \Phi^{(j)}(\mathbf{x}), \tag{29b}$$

$$\Theta(\mathbf{x}) = \sum_{j=0}^{+\infty} \varepsilon^j \, \Theta^{(j)}(\mathbf{x}). \tag{29c}$$

A series of recursive differential problems in terms of $U_k^{(j)}$, $\Phi^{(j)}$, and $\Theta^{(j)}$ can be obtained substituting expansions (29) into (27). Truncating expansions (29) at the zeroth order, the governing equations of the equivalent first-order (Cauchy) homogenized continuum can be obtained. They can be written in terms of the components $C_{iq_1pq_2}$, $\beta_{q_1q_2}$, $K_{q_1q_2}$, $e_{iq_1q_2}$, $\alpha_{iq_1}$, and $\gamma_{q_1}$ of the overall constitutive tensors as

$$C_{iq_1pq_2} \frac{\partial^2 U_p(\mathbf{x})}{\partial x_{q_1} \partial x_{q_2}} + e_{iq_1q_2} \frac{\partial^2 \Phi(\mathbf{x})}{\partial x_{q_1} \partial x_{q_2}} - \alpha_{iq_1} \frac{\partial \Theta(\mathbf{x})}{\partial x_{q_1}} + b_i(\mathbf{x}) = 0, \tag{30a}$$

$$e_{pq_1q_2} \frac{\partial^2 U_p(\mathbf{x})}{\partial x_{q_1} \partial x_{q_2}} - \beta_{q_1q_2} \frac{\partial^2 \Phi(\mathbf{x})}{\partial x_{q_1} \partial x_{q_2}} + \gamma_{q_1} \frac{\partial \Theta(\mathbf{x})}{\partial x_{q_1}} - \rho_e(\mathbf{x}) = 0, \tag{30b}$$

$$K_{q_1q_2} \frac{\partial^2 \Theta(\mathbf{x})}{\partial x_{q_1} \partial x_{q_2}} + r(\mathbf{x}) = 0. \tag{30c}$$

As detailed in (Fantoni et al., 2017) the relations between the components of tensors $\mathbf{n}^{(2)}$, $\mathbf{w}^{(2)}$, $\mathbf{m}^{(2)}$, $\tilde{\mathbf{n}}^{(2)}$, and $\tilde{\mathbf{w}}^{(2)}$, as expressed in equations (23) and (26), and the components of the corresponding overall constitutive tensors $\mathbb{C}, \boldsymbol{\beta}, \mathbf{K}$, and $\mathbf{e}$, can be expressed as

$$n^{(2)}_{ipq_1q_2} = \frac{1}{2} \left( C_{pq_1iq_2} + C_{pq_2iq_1} \right), \quad w^{(2)}_{q_1q_2} = \beta_{q_1q_2}, \quad m^{(2)}_{q_1q_2} = K_{q_1q_2}, \quad \tilde{n}^{(2)}_{iq_1q_2} = \tilde{w}^{(2)}_{iq_1q_2} = e_{iq_1q_2}.$$

Symmetry and positive definiteness of $\mathbf{n}^{(2)}$, $\mathbf{w}^{(2)}$, and $\mathbf{m}^{(2)}$, and the equality between the components of tensor $\tilde{\mathbf{n}}^{(2)}$ and $\tilde{\mathbf{w}}^{(2)}$ are also proved in detail in (Fantoni et al., 2017).

# 4 Application of asymptotic homogenization technique to microstructured thermo-piezoelectric bender actuators

In the present *Section* deflection of multilayered microstructured bender actuators is investigated in relation to some geometric parameters. In particular, unimorph and bimorph thermo-piezoelectric benders are considered (see figure 2); the first one made of an active PZT layer and a composite layer, while the second one made of two PZT laminae and a central composite stratum sandwiched among the two active layers. In both cases, the composite layer is characterized by a periodic microstructure, with PZT inclusions immersed in a polymeric matrix. Electrodes are present on the upper and lower surfaces of each active PZT lamina, as illustrated in figure 2 with a thick black line. As depicted in figure 2, benders are electrically charged by means of an imposed voltage $\Delta\Phi$ between each couple of electrodes. Furthermore, in order to exploit the pyroelectric properties of the PZT, benders are considered subject to a relative temperature gradient $\Delta\Theta$ between the extrados of the actuators, characterized by the macro relative temperature $\Theta_{ext}$, and the intrados of the devices, which is at relative temperature $\Theta_{int}$. In *Section 4.1* constitutive tensors relative to the materials constituting the benders at the microscale are provided, together with microscopic constitutive equations describing plane strain and plane stress conditions. *Section 4.2* is devoted to perform the asymptotic homogenization, as previously described in *Section 3*, of the microstructured composite layer of the benders, where inclusions with different topologies are considered. Therefore, investigated actuators remain heterogeneous at the macroscale. Perturbation functions are determined as solutions of non homogeneous cell problems at the order $\varepsilon^{-1}$ and the overall constitutive tensors of the first-order (Cauchy) homogenized medium are computed for the case at hand.

Finally, in *Section 4.3* deflection of unimorph and bimorph benders is investigated under different loading conditions, in order to analyze the role of the microstructure on the stiffness of the whole actuator.

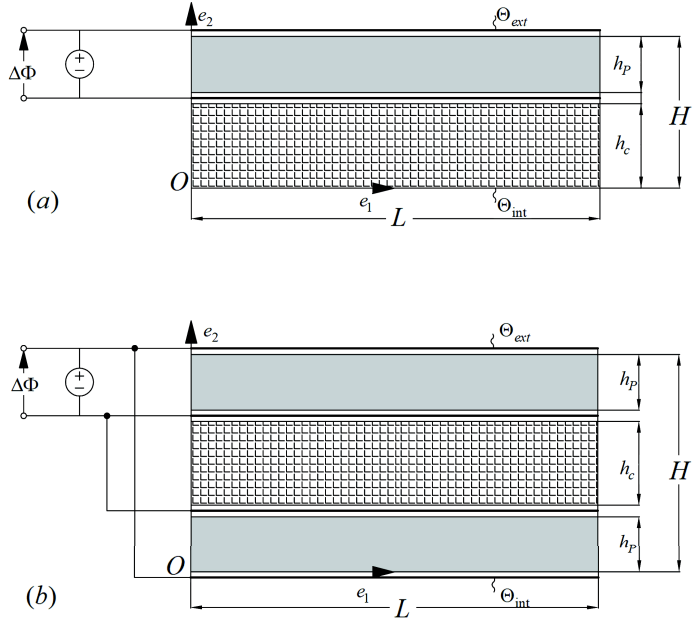

Figure 2: *Bender actuators with height $H$ and length $L$: unimorph (a) and bimorph (b). PZT laminae, with thickness $h_P$, are represented by grey uniform color, while composite layer, with thickness $h_c$, is depicted with a checked background. Electrods are indicated with a thick black line. Benders are loaded by voltage $\Delta\Phi$ and relative temperature gradient $\Delta\Theta = \Theta_{ext} - \Theta_{int}$.*

## 4.1 Microscopic constitutive tensors and in plane description

Consider a unimorph bender as the one depicted in figure 2-(a), formed by a pyroelectric layer and a composite layer. The pyroelectric layer is made by Lead Zirconate Titanate (PZT-5H), which has marked piezoelectric and pyroelectric properties. Such a material is considered polarized in the $x_2$ direction and it is characterized by the following constitutive tensors (Guo et al., 2003; Kommepalli et al., 2010; Malmonge et al., 2003; Umemiya et al., 2006; Yang, 2004), accordingly to the notation detailed in Appendix A, where symbol $|_{x_i}$ refers to the $i^{th}$ polarization direction

$$\mathbb{C}_{PZT|x_2} = \begin{pmatrix} 12.6 & 8.41 & 7.95 & 0 & 0 & 0 \\ 8.41 & 11.7 & 8.41 & 0 & 0 & 0 \\ 7.95 & 8.41 & 12.6 & 0 & 0 & 0 \\ 0 & 0 & 0 & 2 \cdot 2.3 & 0 & 0 \\ 0 & 0 & 0 & 0 & 2 \cdot 2.3 & 0 \\ 0 & 0 & 0 & 0 & 0 & 2 \cdot 2.3 \end{pmatrix} 10^{10} \frac{\text{N}}{\text{m}^2},$$

$$\boldsymbol{\beta}_{PZT|x_2} = \begin{pmatrix} 1.505 & 0 & 0 \\ 0 & 1.302 & 0 \\ 0 & 0 & 1.505 \end{pmatrix} 10^{-8} \frac{\text{C}}{\text{V m}}, \quad \boldsymbol{K}_{PZT|x_2} = \begin{pmatrix} 1.5 & 0 & 0 \\ 0 & 1.5 & 0 \\ 0 & 0 & 1.5 \end{pmatrix} \frac{\text{W}}{\text{m K}},$$

$$\tilde{\boldsymbol{e}}_{PZT|x_2} = \begin{pmatrix} 0 & 0 & 0 & \sqrt{2} \cdot 17 & 0 & 0 \\ -6.5 & 23.3 & 0 & 0 & 0 & 0 \\ 0 & 0 & 0 & 0 & 0 & 0 \end{pmatrix} \frac{\text{C}}{\text{m}^2},$$

$$\boldsymbol{\alpha}_{PZT|x_2} = \begin{pmatrix} 1.71 \\ 1.71 \\ 1.71 \\ 0 \\ 0 \\ 0 \end{pmatrix} 10^6 \frac{\text{N}}{\text{m}^2 \text{K}}, \quad \boldsymbol{\gamma}_{PZT|x_2} = \begin{pmatrix} 5 \\ 5 \\ 5 \end{pmatrix} 10^{-4} \frac{\text{C}}{\text{m}^2 K}. \tag{31}$$

The composite layer has a micro-structure where PZT fibers, polarized in the $x_3$ direction, are immersed in a polymeric matrix. Fibers having circular or square/rectangular cross section have been considered, as depicted in figure 3, whose constitutive tensors are (Yang, 2004)

$$\mathbb{C}_{PZT|x_3} = \begin{pmatrix} 12.6 & 7.95 & 8.41 & 0 & 0 & 0 \\ 7.95 & 12.6 & 8.41 & 0 & 0 & 0 \\ 8.41 & 8.41 & 11.7 & 0 & 0 & 0 \\ 0 & 0 & 0 & 2 \cdot 2.3 & 0 & 0 \\ 0 & 0 & 0 & 0 & 2 \cdot 2.3 & 0 \\ 0 & 0 & 0 & 0 & 0 & 2 \cdot 2.3 \end{pmatrix} 10^{10} \frac{\text{N}}{\text{m}^2},$$

$$\boldsymbol{\beta}_{PZT|x_3} = \begin{pmatrix} 1.505 & 0 & 0 \\ 0 & 1.505 & 0 \\ 0 & 0 & 1.302 \end{pmatrix} 10^{-8} \frac{\text{C}}{\text{V m}},$$

$$\tilde{\boldsymbol{e}}_{PZT|x_3} = \begin{pmatrix} 0 & 0 & 0 & 0 & \sqrt{2} \cdot 17 & 0 \\ 0 & 0 & 0 & \sqrt{2} \cdot 17 & 0 & 0 \\ -6.5 & -6.5 & 23.3 & 0 & 0 & 0 \end{pmatrix} \frac{\text{C}}{\text{m}^2},$$

$$\tag{32}$$

Tensors $\boldsymbol{K}_{PZT|x_3}, \boldsymbol{\alpha}_{PZT|x_3}$ and $\boldsymbol{\gamma}_{PZT|x_3}$ are the same as the ones detailed in equation (31).

Polymeric matrix is considered made of low density polyethylene (LDP), which has negligible piezoelectric and pyroelectric properties (i.e. $\boldsymbol{e}_{LDP} = \boldsymbol{0}$ and $\boldsymbol{\gamma}_{LDP} = \boldsymbol{0}$) and it is characterized by the following constitutive tensors (Peacock, 2000)

$$\mathbb{C}_{LDP} = \begin{pmatrix} 3.69 & 2.46 & 2.46 & 0 & 0 & 0 \\ 2.46 & 3.69 & 2.46 & 0 & 0 & 0 \\ 8.41 & 2.46 & 3.69 & 0 & 0 & 0 \\ 0 & 0 & 0 & 2 \cdot 0.61 & 0 & 0 \\ 0 & 0 & 0 & 0 & 2 \cdot 0.61 & 0 \\ 0 & 0 & 0 & 0 & 0 & 2 \cdot 0.61 \end{pmatrix} 10^8 \frac{\text{N}}{\text{m}^2},$$

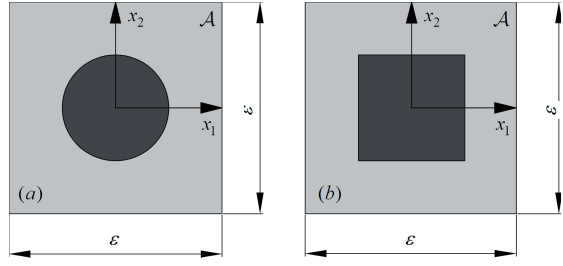

Figure 3: *Periodic cell $\mathcal{A}$ with PZT inclusions (dark grey) immersed in a low density polyethylene matrix (light grey):(a) circular inclusion, (b) square inclusion*

$$\boldsymbol{\beta}_{LDP} = \begin{pmatrix} 2.04 & 0 & 0 \\ 0 & 2.04 & 0 \\ 0 & 0 & 2.04 \end{pmatrix} 10^{-11} \frac{\text{C}}{\text{V m}}, \quad \boldsymbol{K}_{LDP} = \begin{pmatrix} 0.42 & 0 & 0 \\ 0 & 0.42 & 0 \\ 0 & 0 & 0.42 \end{pmatrix} \frac{\text{W}}{\text{m K}},$$

$$\boldsymbol{\alpha}_{LDP} = \begin{pmatrix} 1.03 \\ 1.03 \\ 1.03 \\ 0 \\ 0 \\ 0 \end{pmatrix} 10^5 \frac{\text{N}}{\text{m}^2 \text{K}}, \tag{33}$$

Bimorph bender (see figure 2-(b)) is formed by a central layer made of low density polyethylene and PZT fibers polarized in the $x_3$ direction whose constitutive tensors are described respectively in equations (32) and (33). On the upper and lower surfaces of the composite layer there are two PZT layers polarized in the $x_2$ direction, having constitutive properties described in equation (31). Unimorph and bimorph benders as the ones previously described have been analyzed considering plane strain and plane stress conditions. In the case of plane strain conditions, the gradients of all cinematic variables in the $x_3$ direction have vanishing values, namely

$$\varepsilon_{33} = \varepsilon_{13} = \varepsilon_{23} = 0, \quad E_3 = 0, \quad \theta_{,3} = 0.$$

In such conditions, components of the stress tensors are expressed as

$$\begin{aligned}
\sigma_{11} &= C_{1111}\varepsilon_{11} + C_{1122}\varepsilon_{22} - \alpha_{11}\theta, & \sigma_{22} &= C_{2222}\varepsilon_{22} + C_{2211}\varepsilon_{11} - \alpha_{22}\theta, \\
\sigma_{33} &= C_{1133}\varepsilon_{11} + C_{2233}\varepsilon_{22} - \alpha_{33}\theta, & \sigma_{12} &= 2\,C_{1212}\varepsilon_{12}, \\
\sigma_{13} &= e_{131}\phi_{,1}, & \sigma_{23} &= e_{232}\phi_{,2}.
\end{aligned} \tag{34}$$

Electric displacement $\mathbf{D}$ results

$$D_1 = -\beta_{11}\phi_{,1} + \gamma_1\theta, \quad D_2 = -\beta_{22}\phi_{,2} + \gamma_2\theta, \quad D_3 = \tilde{e}_{311}\varepsilon_{11} + \tilde{e}_{322}\varepsilon_{22} + \gamma_3\theta, \tag{35}$$

while heat flux $q$ has components

$$q_1 = -K_{11}\theta_{,1}, \quad q_2 = -K_{22}\theta_{,2}, \quad q_3 = 0. \tag{36}$$

If plane stress condition is considered, all fluxes in the $x_3$ direction have vanishing values

$$\sigma_{33} = \sigma_{13} = \sigma_{23} = 0, \quad D_3 = 0, \quad q_3 = 0.$$

In this second case, condensation of constitutive tensors leads to the following constitutive equations for the stress $\boldsymbol{\sigma}$, the electric displacement $\boldsymbol{D}$, and the heat flux $\boldsymbol{q}$ in the $\{x_1, x_2\}$ plane

$$\begin{aligned}
\sigma_{11} &= \left[ C_{1111} - \frac{C_{1133}(C_{1133} + e_{333}e_{113}/\beta_{33})}{C_{3333} + e_{333}^2/\beta_{33}} + \frac{e_{113}^2}{\beta_{33}} - \frac{e_{113}e_{333}(C_{1133} + e_{333}e_{113}/\beta_{33})}{\beta_{33}(C_{3333} + e_{333}^2/\beta_{33})} \right] \varepsilon_{11} \\
&+ \left[ C_{1122} - C_{1133}\frac{C_{2233} + e_{223}e_{333}/\beta_{33}}{C_{3333} + e_{333}^2/\beta_{33}} + \frac{e_{113}e_{223}}{\beta_{33}} - \frac{e_{113}e_{333}(C_{2233} + e_{223}e_{333}/\beta_{33})}{\beta_{33}(C_{3333} + e_{333}^2/\beta_{33})} \right] \varepsilon_{22}
\end{aligned}$$

$$
\begin{aligned}
\sigma_{22} &= -\left[\alpha_{11} - \frac{C_{1133}(\alpha_{33} - e_{333}\gamma_3/\beta_{33})}{C_{3333} + e_{333}^2/\beta_{33}} - \frac{e_{113}\gamma_3}{\beta_{33}} - \frac{e_{113}e_{333}(\alpha_{33} - e_{333}\gamma_3/\beta_{33})}{\beta_{33}(C_{3333} + e_{333}^2/\beta_{33})}\right]\theta, \\
&= \left[C_{2222} - \frac{C_{2233}(C_{2233} + e_{333}e_{223}/\beta_{33})}{C_{3333} + e_{333}^2/\beta_{33}} + \frac{e_{223}^2}{\beta_{33}} - \frac{e_{223}e_{333}(C_{2233} + e_{333}e_{223}/\beta_{33})}{\beta_{33}(C_{3333} + e_{333}^2/\beta_{33})}\right]\varepsilon_{22} \\
&+ \left[C_{2211} - C_{2233}\frac{C_{1133} + e_{113}e_{333}/\beta_{33}}{C_{3333} + e_{333}^2/\beta_{33}} + \frac{e_{223}e_{113}}{\beta_{33}} - \frac{e_{223}e_{333}(C_{1133} + e_{113}e_{333}/\beta_{33})}{\beta_{33}(C_{3333} + e_{333}^2/\beta_{33})}\right]\varepsilon_{11} \\
&- \left[\alpha_{22} - \frac{C_{2233}(\alpha_{33} - e_{333}\gamma_3/\beta_{33})}{C_{3333} + e_{333}^2/\beta_{33}} - \frac{e_{223}\gamma_3}{\beta_{33}} - \frac{e_{223}e_{333}(\alpha_{33} - e_{333}\gamma_3/\beta_{33})}{\beta_{33}(C_{3333} + e_{333}^2/\beta_{33})}\right]\theta, \\
\sigma_{12} &= 2C_{1212}\varepsilon_{12}, \quad (37\text{a})
\end{aligned}
$$

$$
\begin{aligned}
D_1 &= -\left[\beta_{11} + \frac{e_{131}^2}{C_{1313}}\right]\phi_{,1} + \gamma_1\theta, \\
D_2 &= \left[\beta_{22} + \frac{e_{233}^2}{C_{2323}}\right]\phi_{,2} + \gamma_2\theta, \quad (37\text{b})
\end{aligned}
$$

$$
\begin{aligned}
q_1 &= -K_{11}\theta_{,1}, \\
q_2 &= -K_{22}\theta_{,2}. \quad (37\text{c})
\end{aligned}
$$

## 4.2 Homogenization of benders' microstructured layer: perturbation functions and overall constitutive tensors

Solution of cell problems (9), (11), (13), and (15) allows to derive perturbation functions relative to the order $\varepsilon^{-1}$. Perturbation functions take into account the effects of material inhomogeneities on micro displacements, electric potential, and relative temperature and on coupling among them. Such functions only depend on geometrical and physico-mechanical properties of the material. One considers periodic cells $\mathcal{A}$ as the ones illustrated in figures 3, characterized respectively by a circular and a square inclusion. Figures 4 and 5 show some of the perturbation functions derived on the unit cell $\mathcal{Q}$ from a numerical resolution of cell problems (9), (11), (13), and (15) at the order $\varepsilon^{-1}$ in plane strain conditions trough a finite element discretization. Illustrated perturbation functions refer to the case of a volume fraction $f = 0.25$ for both the circular inclusion (figure 4), and the square one (figure 5), where $f$ is defined as the ratio between the area of the inclusion and the area of the periodic cell. It is evident from figures 4 and 5 that perturbation functions have vanishing mean values over $\mathcal{Q}$ for the imposed normalization condition (8). Furthermore, perturbation functions are $\mathcal{Q}$-periodic and smooth along the unit cell boundaries.

Overall constitutive tensors of the first-order homogenized medium have been computed by means of the closed forms (23) and (26) for the plane stress and plane strain conditions described above. For example, at $f = 0.25$ and for circular inclusion, indicated by apex $^{circ}$, overall constitutive tensors have the following expression in plane strain conditions

$$
\mathbb{C}^{circ} = \begin{pmatrix} 2.45 & 1.80 & 0 \\ 1.80 & 2.45 & 0 \\ 0 & 0 & 2 \cdot 0.30 \end{pmatrix} 10^9 \, \frac{\text{N}}{\text{m}^2}, \quad \boldsymbol{\beta}^{circ} = \begin{pmatrix} 0.61 & 0 \\ 0 & 0.61 \end{pmatrix} 10^{-9} \, \frac{\text{C}}{\text{V\,m}},
$$

$$
\boldsymbol{K}^{circ} = \begin{pmatrix} 0.57 & 0 \\ 0 & 0.57 \end{pmatrix} \frac{\text{W}}{\text{m\,K}}, \quad \boldsymbol{\alpha}^{circ} = \begin{pmatrix} 1.67 \\ 1.67 \\ 0 \end{pmatrix} 10^5 \, \frac{\text{N}}{\text{m}^2\,\text{K}},
$$

$$
\boldsymbol{\gamma}^{circ} = \begin{pmatrix} 0.19 \\ 0.19 \end{pmatrix} 10^{-4} \, \frac{\text{C}}{\text{m}^2\,\text{K}}, \quad (38)
$$

Considering a square inclusion, indicated by apex $^{sq}$, at the same volume fraction $f = 0.25$ and plane strain conditions, overall constitutive tensors result

$$
\mathbb{C}^{sq} = \begin{pmatrix} 5.53 & 2.96 & 0 \\ 2.96 & 5.53 & 0 \\ 0 & 0 & 2 \cdot 0.86 \end{pmatrix} 10^8 \, \frac{\text{N}}{\text{m}^2}, \quad \boldsymbol{\beta}^{sq} = \begin{pmatrix} 0.35 & 0 \\ 0 & 0.35 \end{pmatrix} 10^{-10} \, \frac{\text{C}}{\text{V\,m}},
$$

$$
\boldsymbol{K}^{sq} = \begin{pmatrix} 0.56 & 0 \\ 0 & 0.56 \end{pmatrix} \frac{\text{W}}{\text{m\,K}}, \quad \boldsymbol{\alpha}^{sq} = \begin{pmatrix} 1.05 \\ 1.05 \\ 0 \end{pmatrix} 10^5 \, \frac{\text{N}}{\text{m}^2\,\text{K}},
$$

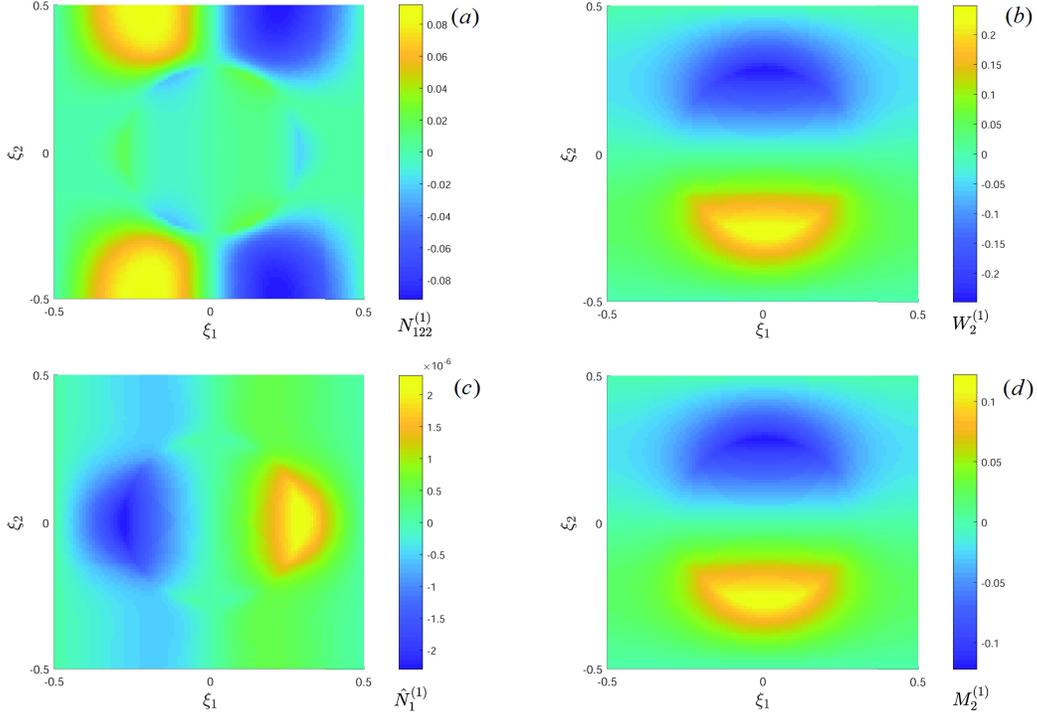

Figure 4: *Some of the perturbation functions obtained by means of numerical resolution of cell problems at the order $\varepsilon^{-1}$ over the unit cell $\mathcal{Q}$ with circular inclusion and $f = 0.25$: (a) $N_{122}^{(1)}$, (b) $W_2^{(1)}$, (c) $\hat{N}_1^{(1)}$, (d) $M_2^{(1)}$*

$$\boldsymbol{\gamma}^{sq} = \begin{pmatrix} 0.49 \\ 0.49 \end{pmatrix} 10^{-6} \frac{\text{C}}{\text{m}^2\,\text{K}}, \tag{39}$$

Figures 6 and 7 show the components of the overall constitutive tensors with respect to volume fraction $f$ for the case of circular and square inclusion, considering plane strain and plane stress conditions. In particular, figures 6 and 7 represent the dimensionless components of the overall tensors, obtained dividing each component of the tensors by the averaged value of the same component relative to the inclusion computed for the plane stress and the plane strain cases. Indicating with $\boldsymbol{B}$ the generic overall tensor, dimensionless overall tensors are therefore defined as:

$$\hat{\boldsymbol{B}} = \frac{2\,\boldsymbol{B}}{\boldsymbol{B}_{PZT|x_3\,pl.strain} + \boldsymbol{B}_{PZT|x_3\,pl.stress}}$$

Figure 6 shows the non vanishing components of the overall constitutive tensors for the case of a circular inclusion whose radius increases monotonically such that $0 \leq f \leq 0.75$.

As shown in figure 6, computed components for the plane strain case are greater than the corresponding ones computed for the plane stress situation for all the overall tensors except in the case of the overall permittivity tensor $\boldsymbol{\beta}$. Figure 7 shows the non vanishing components of the overall constitutive tensors with respect to $f$ for the case of a square inclusion increasing homotetically such that $0 \leq f \leq 0.25$. For values of volume fraction greater than 0.25 a rectangular inclusion has been considered, such that the length along the $x_2$ axis is kept fixed and equal to $0.5\,\varepsilon$, while the length along the $x_1$ axis increases such that $0.25 < f \leq 0.4$. As shown in figure 7, values for the plane strain situation are distinct and greater than the corresponding ones relative to the plane stress case for tensors $\mathbb{C}$, $\boldsymbol{\alpha}$, and $\boldsymbol{\gamma}$. Furthermore it is evident that, for $f > 0.25$, overall tensors stop to be isotropic, diversifying the behavior along the $x_1$ and $x_2$ axes.

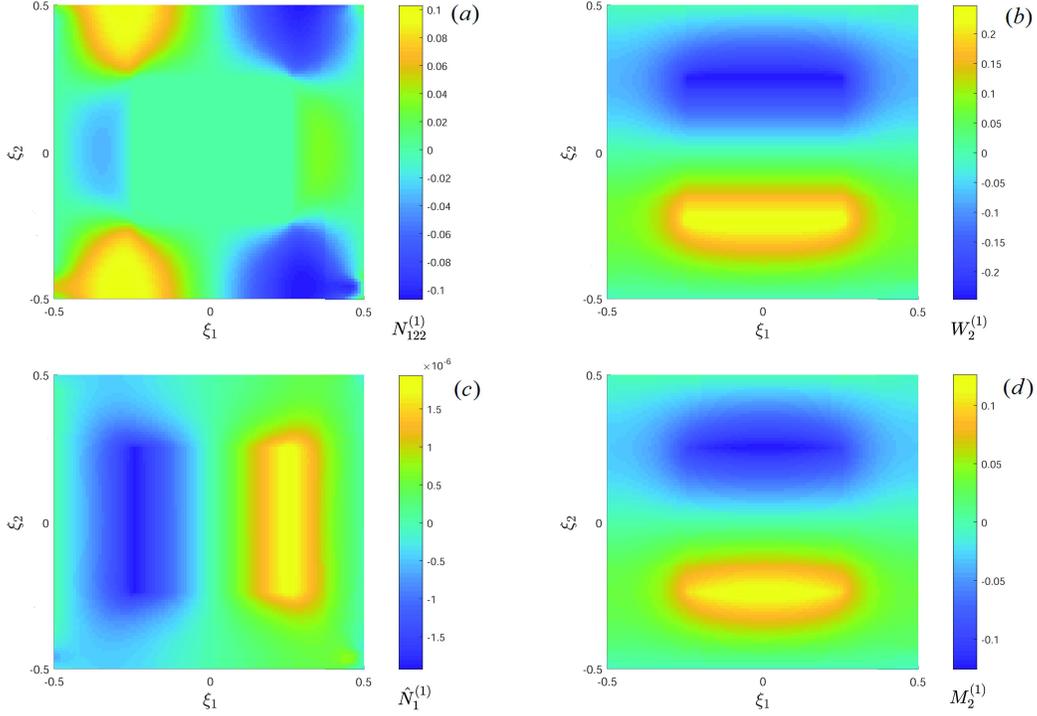

Figure 5: *Some of the perturbation functions obtained by means of numerical resolution of cell problems at the order $\varepsilon^{-1}$ over the unit cell $\mathcal{Q}$ with square inclusion and $f = 0.25$: (a) $N_{122}^{(1)}$, (b) $W_2^{(1)}$, (c) $\hat{N}_1^{(1)}$, (d) $M_2^{(1)}$*

### 4.3 Unimorph and bimorph bender actuators

Deflection of unimorph and bimorph benders as the ones illustrated in figure 2 has been studied in plane strain conditions varying the volume fraction $f$ of the PZT inclusions in the composite layer, considering circular and square/rectangular PZT fibers (see figure 3). Benders are considered fixed at one end trough translational constraints. One defines the following dimensionless macro quantities:

$$x_1^* = \frac{x_1}{L}, \quad x_2^* = \frac{x_2}{H}, \quad U_1^* = \frac{U_1}{L}, \quad U_2^* = \frac{U_2}{L}, \quad \Phi^* = \frac{\Phi\sqrt{\beta_{11_{PZT|x_3}}}}{\sqrt{C_{1111_{PZT|x_3}}}L}, \quad \Theta^* = \frac{\Theta\alpha_{11_{PZT|x_3}}}{C_{11_{PZT|x_3}}}. \qquad (40)$$

Figure 8 shows the deflection of the benders with respect to volume fraction $f$, for three different values of the benders slenderness, namely $H/L = 1/10, 1/20, 1/40$ and a ratio between the thickness of the composite layer and the PZT one equal to $h_c/h_P = 2$. Benders are loaded applying a voltage equal to $\Delta\Phi^* = 5.8 \cdot 10^{(-7)}$ between electrodes, as indicated in figure 2. In particular, dimensionless values of the macro displacement $\tilde{U}_2$ are represented in figure 8, this last defined as the ratio between displacement $U_2$ measured at the free end of benders with coordinates $x_1 = L$ and $x_2 = 0$ and the corresponding value obtained at the same point for $f = 0$, namely $\tilde{U}_2 = U_{2|f}/U_{2|f=0}$. As shown in figures 8-(a) and 8-(c) the deflection of the unimorph bender is not influenced by the slenderness, increasing monotonically as $f$ raises, reaching a maximum value equal to 6.6 for circular inclusion and equal to 3.2 for square inclusion. On the contrary, the behavior of the bimorph bender is strongly influenced by the slenderness, increasing the deflection as the slenderness increases. For the case of the circular inclusion, values of $\tilde{U}_2$ reach their maximum at $f = 0.5$ for each value of the ratio $H/L$ and, in particular, for $H/L = 1/40$, the dimensionless displacement $\tilde{U}_2$ becomes less than 1 for $f > 0.5$, meaning that the presence of the PZT fibers in the composite layer lowers the bender deflection with respect to the case without fibers. For the square inclusion case, deflection of the bimorph

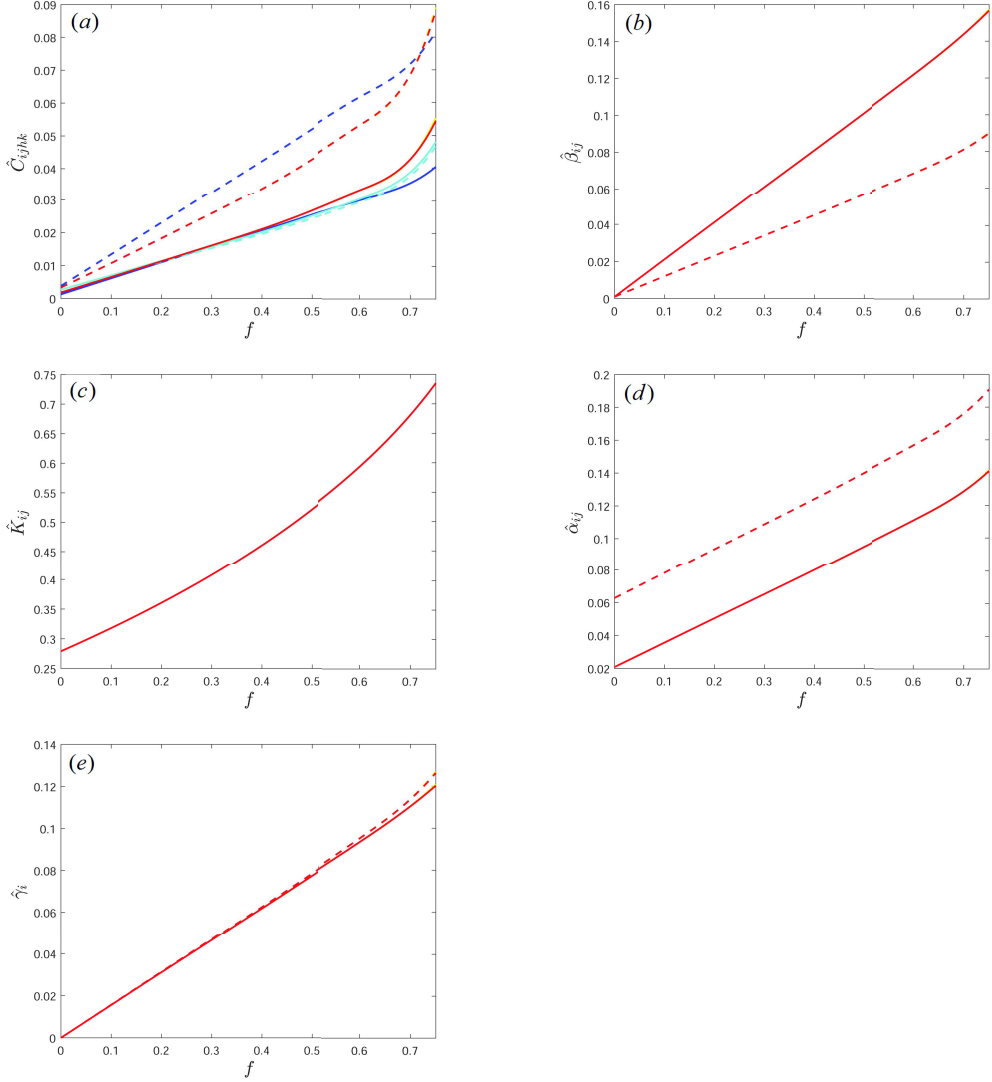

Figure 6: *Dimensionless components of the overall constitutive tensors vs volume fraction $f$, with $0 \leq f \leq 0.75$ and circular inclusion for plane stress (continuous line) and plane strain (dashed line) conditions. (a) $\hat{C}_{1111} = \hat{C}_{2222}$ (red curve), $\hat{C}_{1122}$ (blue curve), $\hat{C}_{1212}$ (light blue curve), (b) $\hat{\beta}_{11} = \hat{\beta}_{22}$, (c) $\hat{K}_{11} = \hat{K}_{22}$, (d) $\hat{\alpha}_{11} = \hat{\alpha}_{22}$, (e) $\hat{\gamma}_1 = \hat{\gamma}_2$*

bender increases monotonically as $f$ increases for $H/L = 1/10$ and $H/L = 1/20$, while reaches its maximum at $f = 0.3$ and decreases for $f > 0.3$ in the case $H/L = 1/40$. Except for the case of bimorph bender and H/L=1/40, for the same value of volume fraction $f$, values of $\tilde{U}_2$ for circular inclusions are greater than the corresponding ones for square inclusions, meaning that the fibers topology plays a role in characterizing the behavior of the actuator. Furthermore, in all cases here analyzed, values of $\tilde{U}_2$ for the bimorph bender are lower than the unimorph bender ones. This means that the presence of a microstructured composite layer influences the stiffness of the unimorph bender more than it does for the bimorph one.

Figure 9 shows the deflection of the unimorph and bimorph benders with respect to $f$ for $H/L = 1/20$ and three different values of the ratio between the thickness of the composite layer and the PZT one, namely $h_c/h_P = 2, 1, 1/2$. As shown in figure 9 values of $\tilde{U}2$ are greatly influenced by the parameter $h_c/h_P$,

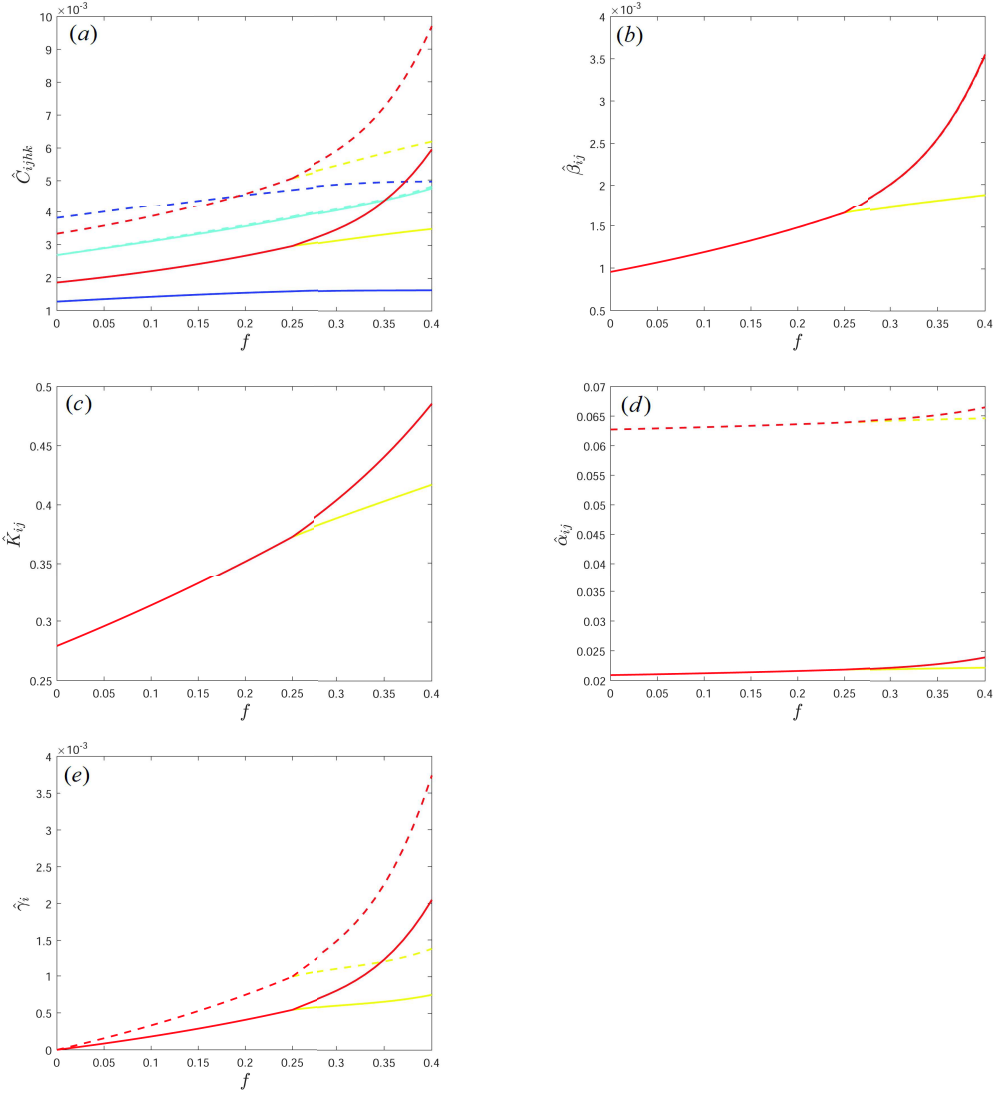

Figure 7: *Dimensionless components of the overall constitutive tensors vs volume fraction $f$, with $0 \leq f \leq 0.4$ and square/rectangular inclusion for plane stress (continuous line) and plane strain (dashed line) conditions. (a) $\hat{C}_{1111}$ (red curve), $\hat{C}_{2222}$ (yellow curve), $\hat{C}_{1122}$ (blue curve), $\hat{C}_{1212}$ (light blue curve), (b) $\hat{\beta}_{11}$ (red curve), $\hat{\beta}_{22}$ (yellow curve), (c) $\hat{K}_{11}$ (red curve), $\hat{K}_{22}$ (yellow curve), (d) $\hat{\alpha}_{11}$ (red curve), $\hat{\alpha}_{22}$ (yellow curve), (e) $\hat{\gamma}_1$ (red curve), $\hat{\gamma}_2$ (yellow curve)*

increasing as $h_c/h_P$ decreases. Values of dimensionless displacement $\tilde{U}_2$ increase monotonically in all cases except for the case of bimorph bender with circular inclusion and $h_c/h_P = 1$ and $h_c/h_P = 2$ for which $\tilde{U}_2$ reaches its maximum at $f = 0.6$. Once again, values of $\tilde{U}_2$ for the circular inclusion cases are greater than the corresponding ones for square inclusion at the same $f$ and $\tilde{U}_2$ values for the unimorph bender are greater than the ones relative to the bimorph bender, reaching values almost equal to 30 for the case of unimorph bender with circular inclusion and $h_c/h_P = 1/2$.

Furthermore, in order to analyze the influence of the pyroelectric properties of the material on the global behavior of the bender, deflection of unimorph and bimorph benders has been studied by applying, in addition to voltage $\Delta\Phi^*$, a relative temperature gradient between the upper and lowers edges of the actuators equal to $\Delta\Theta^* = \Theta^*_{ext} - \Theta^*_{int}$ (see figure 2). Figure 10 shows the dimensionless displacement $\tilde{U}_2$ at the free end of the benders with respect to volume fraction $f$ for five different values of the ratio $\eta = \Delta\Phi^*/\Delta\Theta^*$, namely $\eta = 1/10, 1/5, 1, 5, 10$.

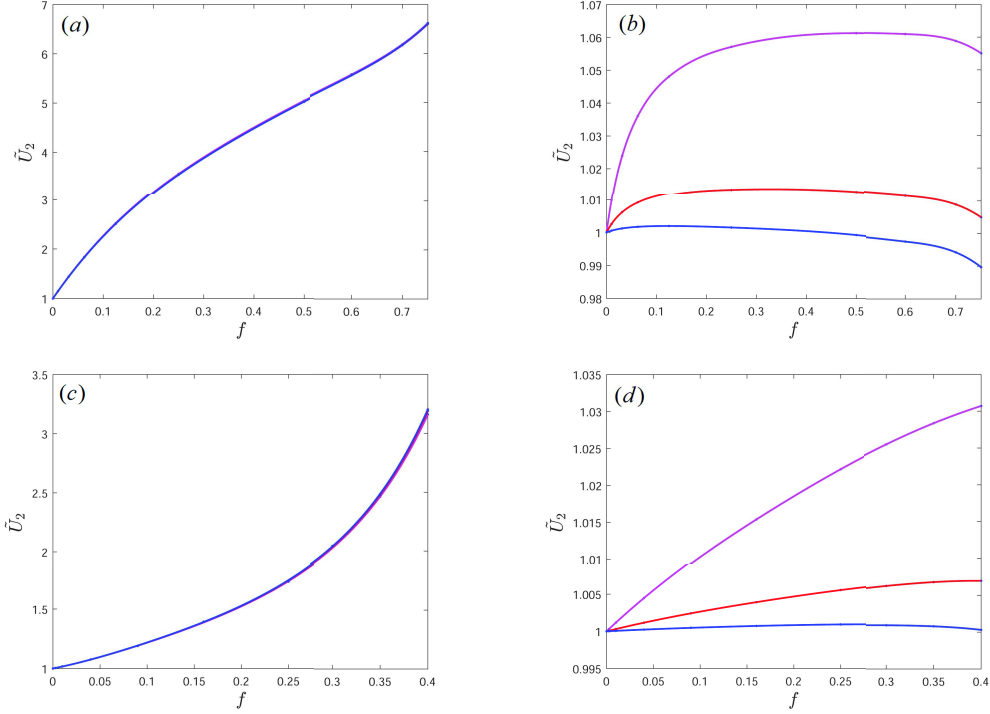

Figure 8: *Dimensionless macro displacement $\tilde{U}_2$ vs volume fraction $f$ for $h_c/h_P = 2$ and three different values of bender slenderness $H/L = 1/10$ (magenta curves), $H/L = 1/20$ (red curves), and $H/L = 1/40$ (blue curves). (a) unimorph bender and circular inclusion, (b) bymorph bender and circular inclusion, (c) unimorph bender and square/rectangular inclusion, (d) bimorph bender and square/rectangular inclusion.*

As shown in figure 10, $\tilde{U}_2$ increases as $\eta$ increases. For values of $\eta \leq 1$, deflection of the benders greatly depends on $\eta$, while this doesn't happen when $\eta > 1$. In fact, obtained curves for $\eta = 5$ and $\eta = 10$ are almost overlapped for all the cases represented in figure 10. Dimensionless displacement $\tilde{U}_2$ increases monotonically with $f$, except for the case of bimorph bender and circular inclusion represented in figure 10-(b). Furthermore, for the case of unimorph bender and circular inclusion, the heterogeneity of the material at the macroscale is responsible of the presence of surface forces which act at the interface between the homogenized composite layer and the PZT one whose values are such that the bender manifests a non intuitive behavior for $f \leq 0.125$ when subject to $\Delta\Theta^*$. Such surface forces are generated by the mismatch that exists between the components of the elastic tensor $\mathbb{C}_{PZT|x_2}$ and of the thermal dilatation tensor $\boldsymbol{\alpha}_{PZT|x_2}$ relative to the PZT layer and the corresponding ones of the homogenized composite layer. Namely, if $\Theta^*_{ext} > \Theta^*_{int}$, the bender deflects upward if $f \leq 0.125$, contrarily to what would happen if the material was homogeneous. If one considers the not normalized macro displacement $U_2$, it is evident that at $f = 0.125$ there is an inversion of tendency of $U_2$ curves with respect to $\eta$.

Finally, figures 11 and 12 show the contour plots of the dimensionless macro fields $U_1^*, U_2^*, \Phi^*$ and $\Theta^*$, as expressed in equation (40), for one of the analyzed load cases, namely $\eta = 1$, $H/L = 1/20$, $h_c/h_P = 2$, circular inclusion and $f = 1/4$ for the unimorph and bimorph benders.

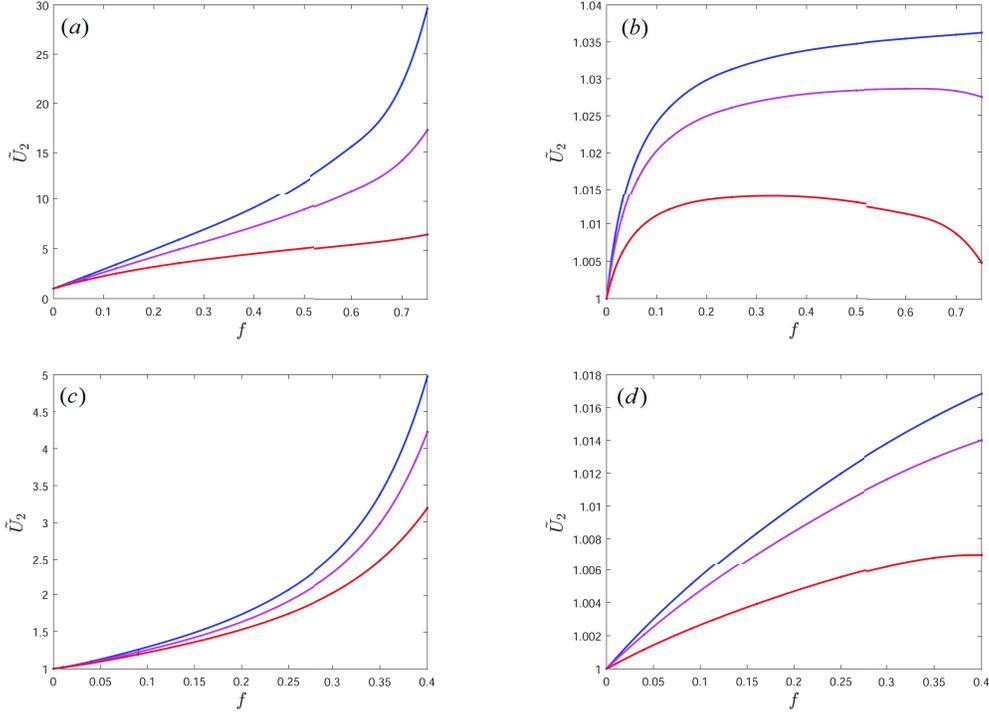

Figure 9: *Dimensionless macro displacement $\tilde{U}_2$ vs volume fraction $f$ for $H/L = 1/20$ and three different values of $h_c/h_P$, namely $h_c/h_P = 2$ (red curves), $h_c/h_P = 1$ (magenta curves), and $h_c/h_P = 1/2$ (blue curves). (a) unimorph bender and circular inclusion, (b) bymorph bender and circular inclusion, (c) unimorph bender and square/rectangular inclusion, (d) bimorph bender and square/rectangular inclusion.*

## 5 Conclusions

Applications of piezoelectric and pyroelectric devices are several, especially as sensing and actuating means. High precision required in the manufacturing process, remarkably in the case of MEMS, leads to the need of accurate preliminary computations. In this regard, determination of proper constitutive relations characterizing the medium at the macroscale could reveal of great importance in order to optimize the performance of the device. The multi-field asymptotic homogenization model obtained in (Fantoni et al., 2017) is here presented as it constitutes a rigorous tool for the characterization at the macroscale of thermo-piezolectric materials with periodic microstructure, avoiding the challenging computational need to model the whole heterogeneous structure. Down-scaling relations are provided; they associate the microscopic displacement, electric potential and relative temperature to the corresponding macroscopic fields and their coupling by means of perturbation functions. These lasts are obtained trough the resolution of non homogeneous recursive cell problems over the unit cell $\mathcal{Q}$. Such functions reflect the effect of the microstructural heterogeneity on the micro fields and on the coupling between them. They result to be $\mathcal{Q}$-periodic, with a vanishing mean value over $\mathcal{Q}$ for the imposed normalization condition. Field equations of the homogenized first-order (Cauchy) medium equivalent to the heterogeneous thermo-piezoelectric one have been determined truncating at the zeroth order the average field equations of infinite order asymptotically expanded in powers of the microstructural size in terms of the macroscopic fields. Exact expressions of the overall constitutive tensors have been determined for the equivalent first-order homogenized material. The accuracy of the obtained multi-field first-order asymptotic homogenization technique has been assessed in (Fantoni et al., 2017), where the very good agreement obtained between the homogenized solution and the numerical one relative to the heterogeneous model subject to harmonic volume forces confirm the validity of the proposed method.

In the present study unimorph and bimorph pyroelectric bending actuators are analyzed in plane stress and plane strain conditions. They consist of active laminae made of PZT with an in plane polarization and a microstructured composite layer with periodic microstructure where PZT fibers, polarized in the out of plane direction, are immersed in a polymeric passive matrix. The previously described multi-field asymptotic technique, developed for thermo-piezoelectric materials, has been exploited to derive the constitutive law of the composite layer at the macroscale, considering different topologies of the PZT inclusions. Deflection of the benders has been investigated in relation to some geometrical features in order to study the influence of the microstructure on the overall stiffness of the actuator. From the analyzed cases it resulted that in the case of composite layer with circular inclusions the benders, loaded by an imposed voltage and/or relative temperature gradient, deflect more than for the case of square/rectangular fibers at the same values of inclusion volume fraction, both for the unimorph and bimorph case. Circular inclusions are therefore preferable to square ones in order to increase the performance of the devices. Furthermore, in all the investigated cases unimorph benders deflect more than bimorph devices do for the same inclusion topology and at the same values of volume fraction of the inclusions, meaning that the presence of a microstructured composite layer influences the stiffness of the unimorph actuators more than it does for the bimorph devices. In particular, in the presence of a microstructured composite layer unimorph benders increase their performance much more than the bimorph counterparts do with respect to the case of a passive layer without pyroelectric inclusions. Concluding, the evaluation of the overall thermo-piezoelectric properties of pyroelectric devices through the multi-field asymptotic homogenization approach illustrated in the paper can represent an important issue in order to improve the efficient design and manufacturing of such systems. Nevertheless, the proposed first-order asymptotic homogenization model can be extended to higher orders to properly describe non local phenomena or, equivalently, high gradients of stresses, deformations, electric potential, relative temperature, and volume forces. Alternatively, in order to derive constitutive relations of equivalent higher order continua, nonlocal higher order homogenization techniques involving characteristic length scale associated to the microstructural effects, can be conveniently deployed, but these topics are left for further research.

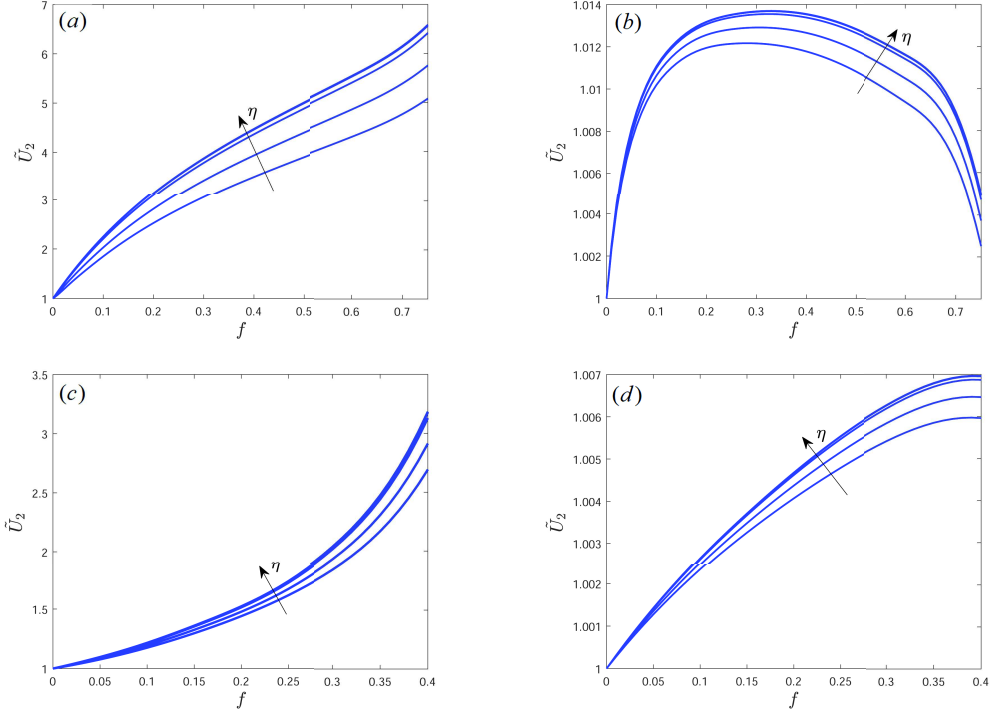

Figure 10: *Dimensionless macro displacement $\tilde{U}_2$ vs volume fraction $f$ for five different values of $\eta = \Delta\Phi^*/\Delta\Theta^*$, namely $\eta = 1/10, 1/5, 1, 5, 10$. Arrows in the graphs indicate the direction of increasing $\eta$. (a) Unimorph bender and circular inclusion, (b) bymorph bender and circular inclusion, (c) unimorph bender and square/rectangular inclusion, (d) bimorph bender and square/rectangular inclusion.*

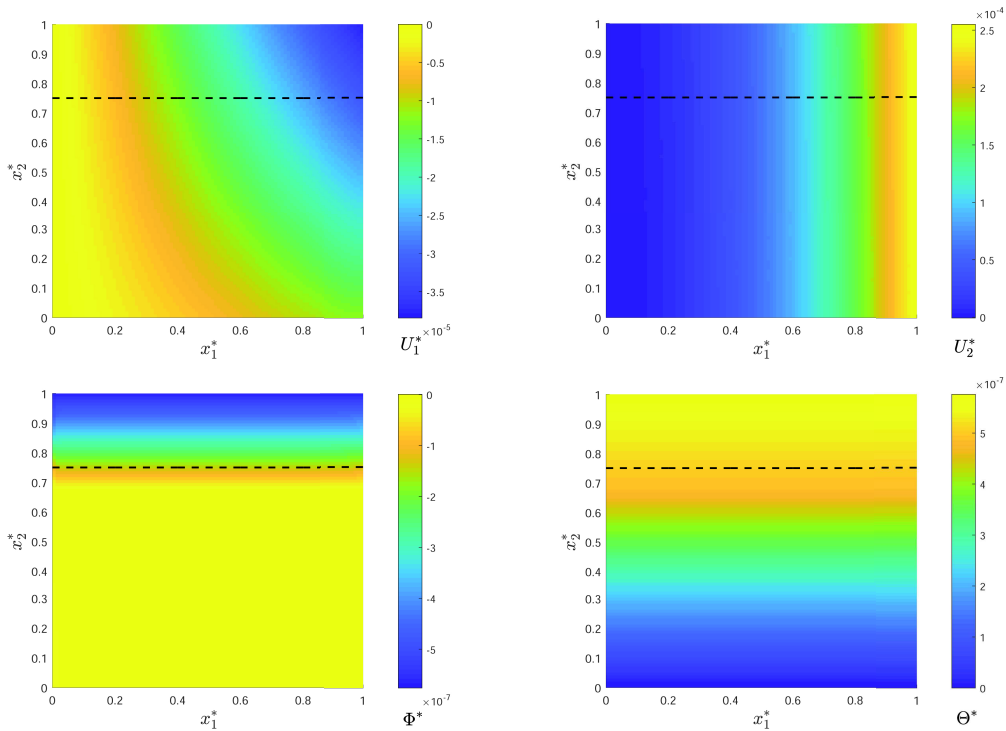

Figure 11: *Contour plots of the macro fields relative to the unimorph bender with circular fibers and $f = 1/4$ subject to $\Delta\Phi^*$ and $\Delta\Theta^*$ such that $\eta = \Delta\Phi^*/\Delta\Theta^* = 1$. Black dashed line indicates the interface between the PZT layer and the composite one. (a) $U_1^*$, (b) $U_2^*$, (c) $\Phi^*$, (d) $\Theta^*$.*

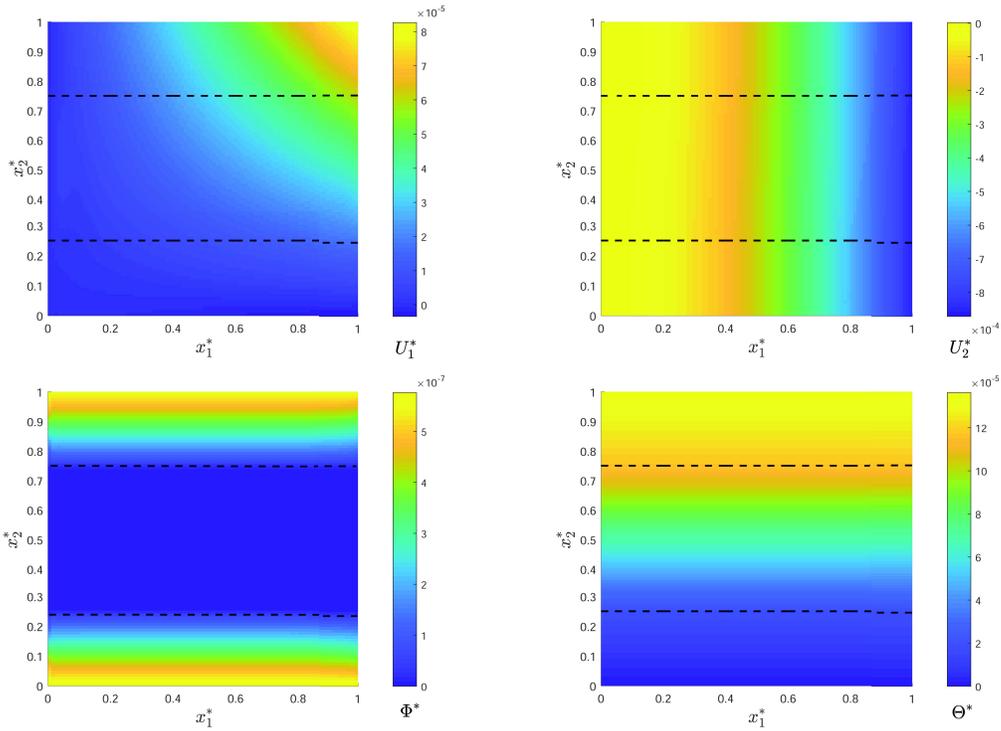

Figure 12: *Contour plots of the macro fields relative to the bimorph bender with circular fibers and $f = 1/4$ subject to $\Delta\Phi^*$ and $\Delta\Theta^*$ such that $\eta = \Delta\Phi^*/\Delta\Theta^* = 1$. Black dashed lines indicate the interfaces between the PZT layers and the composite one. (a) $U_1^*$, (b) $U_2^*$, (c) $\Phi^*$, (d) $\Theta^*$.*

## A  Tensorial notation for constitutive tensors

The expression of the linear constitutive equations (3a)-(3c) for the thermo-piezoelectric material is

$$\begin{aligned}
\sigma_{ij} &= C^m_{ijkl}\, u_{k,l} + e^m_{ijk}\, \phi_{,k} - \alpha^m_{ij}\, \theta, \\
D_i &= e^m_{kli}\, u_{k,l} - \beta^m_{il}\, \phi_{,l} + \gamma^m_i\, \theta, \\
q_i &= -K^m_{ij}\, \theta_{,j},
\end{aligned} \qquad (41)$$

The rigorous form of constitutive equations (41) in a tensorial fashion is (Mehrabadi and Cowin, 1990)

$$\begin{pmatrix} \sigma_{11} \\ \sigma_{22} \\ \sigma_{33} \\ \sqrt{2}\,\sigma_{23} \\ \sqrt{2}\,\sigma_{13} \\ \sqrt{2}\,\sigma_{12} \end{pmatrix} = \begin{pmatrix} C^m_{1111} & C^m_{1122} & C^m_{1133} & \sqrt{2}\,C^m_{1123} & \sqrt{2}\,C^m_{1113} & \sqrt{2}\,C^m_{1112} \\ C^m_{2211} & C^m_{2222} & C^m_{2233} & \sqrt{2}\,C^m_{2223} & \sqrt{2}\,C^m_{2213} & \sqrt{2}\,C^m_{2212} \\ C^m_{3311} & C^m_{3322} & C^m_{3333} & \sqrt{2}\,C^m_{3323} & \sqrt{2}\,C^m_{3313} & \sqrt{2}\,C^m_{3312} \\ \sqrt{2}\,C^m_{2311} & \sqrt{2}\,C^m_{2322} & \sqrt{2}\,C^m_{2333} & 2\,C^m_{2323} & 2\,C^m_{2313} & 2\,C^m_{2312} \\ \sqrt{2}\,C^m_{1311} & \sqrt{2}\,C^m_{1322} & \sqrt{2}\,C^m_{1333} & 2\,C^m_{1323} & 2\,C^m_{1313} & 2\,C^m_{1312} \\ \sqrt{2}\,C^m_{1211} & \sqrt{2}\,C^m_{1222} & \sqrt{2}\,C^m_{1233} & 2\,C^m_{1223} & 2\,C^m_{1213} & 2\,C^m_{1212} \end{pmatrix} \begin{pmatrix} u_{1,1} \\ u_{2,2} \\ u_{3,3} \\ \frac{\sqrt{2}}{2}\,(u_{2,3}+u_{3,2}) \\ \frac{\sqrt{2}}{2}\,(u_{1,3}+u_{3,1}) \\ \frac{\sqrt{2}}{2}\,(u_{1,2}+u_{2,1}) \end{pmatrix} +$$

$$
+ \begin{pmatrix} e_{111}^m & e_{112}^m & e_{113}^m \\ e_{221}^m & e_{222}^m & e_{223}^m \\ e_{331}^m & e_{332}^m & e_{333}^m \\ \sqrt{2}\,e_{231}^m & \sqrt{2}\,e_{232}^m & \sqrt{2}\,e_{233}^m \\ \sqrt{2}\,e_{131}^m & \sqrt{2}\,e_{132}^m & \sqrt{2}\,e_{133}^m \\ \sqrt{2}\,e_{121}^m & \sqrt{2}\,e_{122}^m & \sqrt{2}\,e_{123}^m \end{pmatrix} \begin{pmatrix} \phi_{,1} \\ \phi_{,2} \\ \phi_{,3} \end{pmatrix} - \begin{pmatrix} \alpha_{11}^m \\ \alpha_{22}^m \\ \alpha_{33}^m \\ \sqrt{2}\,\alpha_{23}^m \\ \sqrt{2}\,\alpha_{13}^m \\ \sqrt{2}\,\alpha_{12}^m \end{pmatrix} \theta,
$$

$$
\begin{pmatrix} D_1 \\ D_2 \\ D_3 \end{pmatrix} = \begin{pmatrix} e_{111}^m & e_{221}^m & e_{331}^m & \sqrt{2}\,e_{231}^m & \sqrt{2}\,e_{131}^m & \sqrt{2}\,e_{121}^m \\ e_{112}^m & e_{222}^m & e_{332}^m & \sqrt{2}\,e_{232}^m & \sqrt{2}\,e_{132}^m & \sqrt{2}\,e_{122}^m \\ e_{113}^m & e_{223}^m & e_{333}^m & \sqrt{2}\,e_{233}^m & \sqrt{2}\,e_{133}^m & \sqrt{2}\,e_{123}^m \end{pmatrix} \begin{pmatrix} u_{1,1} \\ u_{2,2} \\ u_{3,3} \\ \frac{\sqrt{2}}{2}(u_{2,3}+u_{3,2}) \\ \frac{\sqrt{2}}{2}(u_{1,3}+u_{3,1}) \\ \frac{\sqrt{2}}{2}(u_{1,2}+u_{2,1}) \end{pmatrix} +
$$

$$
- \begin{pmatrix} \beta_{11}^m & \beta_{12}^m & \beta_{13}^m \\ \beta_{21}^m & \beta_{22}^m & \beta_{23}^m \\ \beta_{31}^m & \beta_{32}^m & \beta_{33}^m \end{pmatrix} \begin{pmatrix} \phi_{,1} \\ \phi_{,2} \\ \phi_{,3} \end{pmatrix} + \begin{pmatrix} \gamma_1^m \\ \gamma_2^m \\ \gamma_3^m \end{pmatrix} \theta,
$$

$$
\begin{pmatrix} q_1 \\ q_2 \\ q_3 \end{pmatrix} = - \begin{pmatrix} K_{11}^m & K_{12}^m & K_{13}^m \\ K_{21}^m & K_{22}^m & K_{23}^m \\ K_{31}^m & K_{32}^m & K_{33}^m \end{pmatrix} \begin{pmatrix} \theta_{,1} \\ \theta_{,2} \\ \theta_{,3} \end{pmatrix}. \tag{42}
$$